     \documentclass[preprint]{aastex}

      \usepackage{epsfig}

\newcommand{\ea}{et al.}
\newcommand{\lta}{\lesssim}
\newcommand{\gta}{\gtrsim}

\newcommand{\kms}{\>{\rm km}\,{\rm s}^{-1}}
\newcommand{\pc}{\>{\rm pc}}
\newcommand{\kpc}{\>{\rm kpc}}

\newcommand{\mum}{\>{\mu {\rm m}}}

\newcommand{\myr}{\>{\rm Myr}}

\newcommand{\lsun}{\>{\rm L_{\odot}}}

\newcommand{\as}{^{\prime\prime}}

\newcommand{\bdm}{\begin{displaymath}}
\newcommand{\edm}{\end{displaymath}}
\newcommand{\beq}{\begin{equation}}
\newcommand{\eeq}{\end{equation}}
\newcommand{\bit}{\begin{itemize}}
\newcommand{\eit}{\end{itemize}}
\newcommand{\ben}{\begin{enumerate}}
\newcommand{\een}{\end{enumerate}}
\newcommand{\bfi}{\begin{figure}[htb]}
\newcommand{\bpfi}{\begin{figure}[p]}

\newcommand{\mfir}{$M_{\rm FIR}$}

\newcommand{\mi}{$M_I$}
\newcommand{\mic}{$M_I^C$}
\newcommand{\mappic}{$m_I^C$}

\newcommand{\mbg}{$M_B^G$}
\newcommand{\mappbg}{$m_B^G$}
\newcommand{\mfirg}{$M_{\rm FIR}^G$}
\newcommand{\mucd}{$\mu_I^0$}

\newcommand{\ttim}{{\it TinyTim}}
\newcommand{\ish}{{\it ISHAPE}}
\newcommand{\re}{$r_{\rm e}$}

\shorttitle{Sizes of Nuclear Star Clusters}
\shortauthors{B\"oker \ea }

\begin{document}


%%%%%%%%%%%%%%%
% Useful definitions during manuscript preparation
%%%%%%%%%%%%%%%

% Uncomment this to ignore figures during LaTeX compilation
% \def\epsfbox#1{}

% Macros to mark places where numbers remain to be added (\xx), where something
% remains to be done (\tbd), or where a comment is placed for the co-authors
% (\com).

\def\xx{$^{[xx]}$}

\def\tbd#1{{\baselineskip=9pt\medskip\hrule{\small\tt #1}
\smallskip\hrule\medskip}}

\def\com#1{{\baselineskip=9pt\medskip\hrule{\small\sl #1}
\smallskip\hrule\medskip}}

\title{A {\it Hubble Space Telescope} Census of Nuclear Star Clusters in 
Late-Type Spiral Galaxies: II. Cluster Sizes and Structural Parameter Correlations}


\author{Torsten B\"oker,\altaffilmark{1} Marc Sarzi,\altaffilmark{2}}
%\affil{Space Telescope Science Institute, 3700 San Martin Drive, 
%Baltimore, MD 21218, U.S.A.}
%\email{boeker@stsci.edu}

%\author{Marc Sarzi\altaffilmark{3}}
%\affil{University of Oxford, Keble Rd., Oxford OX1 3RH, England}
%\email{sarzi@astro.ox.ac.uk}

\author{Dean E. McLaughlin,\altaffilmark{3} Roeland P. van der Marel,\altaffilmark{3}}
%\affil{Space Telescope Science Institute, 3700 San Martin Drive, 
%Baltimore, MD 21218, U.S.A.}
%\email{deanm@stsci.edu, marel@stsci.edu}

\author{Hans-Walter Rix,\altaffilmark{4} Luis C. Ho,\altaffilmark{5} \& Joseph C. Shields\altaffilmark{6}}
%\affil{Max-Planck-Institut f\"ur Astronomie, K\"onigsstuhl 17, 
%D-69117 Heidelberg, Germany }
%\email{rix@mpia-hd.mpg.de}

%\author{Luis C. Ho\altaffilmark{5}}
%\affil{Observatories of the Carnegie Institution of Washington, 
%        813 Santa Barbara Street, Pasadena, CA 91101-1292, U.S.A.}
%\email{lho@ociw.edu}

%\author{Joseph C. Shields\altaffilmark{6}}
%\affil{Ohio University, Department of Physics and Astronomy, 
%        Clippinger Research Laboratories, 251B, Athens, OH 45701-2979}
%\email{shields@helios.phy.ohiou.edu}


\altaffiltext{1}{Astrophysics Division, RSSD, European Space Agency, ESTEC, 
      NL-2200 AG Noordwijk, The Netherlands}
\altaffiltext{2}{University of Oxford, Keble Rd., Oxford OX1 3RH, England}
\altaffiltext{3}{Space Telescope Science Institute, 3700 San Martin Drive, 
	Baltimore, MD 21218}
%\altaffiltext{2}{On assignment from the Space Telescope Division of
%	the European Space Agency}
\altaffiltext{4}{Max-Planck-Institut f\"ur Astronomie, K\"onigsstuhl 17, 
	D-69117 Heidelberg, Germany}
\altaffiltext{5}{Observatories of the Carnegie Institution of Washington, 
        813 Santa Barbara Street, Pasadena, CA 91101-1292}
\altaffiltext{6}{Ohio University, Department of Physics and Astronomy, 
        Clippinger Research Laboratories, 251B, Athens, OH 45701-2979}


%%%%%%%%%%%%%%%%%%%%%%%%%%%%%%%%%%%%%%%%%%%%%%%%%%%%%%%%%%%%%%%%%%%%%%
\begin{abstract}
We investigate the structural properties of nuclear star clusters in 
late-type spiral galaxies. More specifically, we fit 
analytical models to {\it Hubble Space Telescope} images of 39
nuclear clusters in order to determine their effective radii after 
correction for the instrumental point spread function. We use the results
of this analysis to compare the luminosities and sizes of nuclear
star clusters to those of other ellipsoidal stellar systems, in
particular the Milky Way globular clusters. Our nuclear clusters have a
median effective radius of $\overline{r_{\rm e}}=3.5$ pc, with 50\% of the
sample falling between $2.4\ {\rm pc}\le r_{\rm e}\le 5.0\ {\rm pc}$. This
narrow size distribution
is statistically indistinguishable from that of Galactic globular clusters,
even though the nuclear clusters are on average 4 magnitudes brighter than the
old globulars. We discuss some possible interpretations of this result.
From a comparison of nuclear cluster luminosities with various
properties of their host galaxies, we confirm that more luminous 
galaxies harbor more luminous nuclear clusters. It remains  unclear whether
this correlation mainly reflects the influence of galaxy size,
mass, and/or star formation rate. Since the brighter galaxies in our
sample typically have stellar disks with a higher central surface 
brightness, nuclear cluster luminosity also correlates with this 
property of their hosts. On the other hand, we find no evidence for 
a correlation between the presence of a  nuclear star cluster and the
presence of a large-scale stellar bar.
\end{abstract}
%%%%%%%%%%%%%%%%%%%%%%%%%%%%%%%%%%%%%%%%%%%%%%%%%%%%%%%%%%%%%%%%%%%%%%
%%%%%%%%%%%%%%%%%%%%%%%%%%%%%%%%%%%%%%%%%%%%%%%%%%%%%%%%%%%%%%%%%%%%%%
%%%%%%%%%%%%%%%%%%%%%%%%%%%%%%%%%%%%%%%%%%%%%%%%%%%%%%%%%%%%%%%%%%%%%%
\section{Introduction}
%%%%%%%%%%%%%%%%%%%%%%%%%%%%%%%%%%%%%%%%%%%%%%%%%%%%%%%%%%%%%%%%%%%%%%
%%%%%%%%%%%%%%%%%%%%%%%%%%%%%%%%%%%%%%%%%%%%%%%%%%%%%%%%%%%%%%%%%%%%%%
In most formation scenarios for spiral galaxies, the central bulge is 
the ``trashbin of violent relaxation'' where a dynamically hot stellar 
component has formed either through external potential perturbations 
such as early mergers of proto-galaxies \citep[e.g.][]{car92}, 
or perhaps via internal 
effects such as violent bar instabilities \citep{nor96}. While this
``nature or nurture'' debate is far from settled, the recent discovery
of the so-called $M$-$\sigma$ relation \citep{geb00,fer00} has made it 
clear that the structure of the very nucleus, i.e. the black hole and its
immediate vicinity ($\ll 1\pc$), is closely linked to the overall dynamical 
structure of the stellar bulge.
The latest-type spirals, then, must have lived very sheltered and uneventful
lives, since their central ``trashbin'' is virtually empty. Indeed, these 
disk-dominated, bulge-less galaxies \citep[e.g.][]{mat97} often have slowly
rising 
rotation curves that indicate a nearly homogeneous mass distribution on
scales $\lta 1\kpc$. On these scales, gravity therefore hardly provides
a vector pointing at the center, and it is not obvious that the nucleus
of these galaxies is well-defined and a unique environment.

It therefore came as a surprise that recent observations with the
{\it Hubble Space Telescope (HST)} have shown that the photocenter
of many late-type spiral galaxies is occupied by a compact, luminous
stellar cluster \citep{phi96,car98,mat99}. In a recent paper 
\citep[][hereafter Paper~I]{boe02}, we have shown that at least 75\% of 
spiral galaxies with Hubble types later than Sc harbor such a nuclear star 
cluster. The fraction of nuclear clusters (NCs) appears to be somewhat lower
in  earlier Hubble types, as indicated by the results of \cite{car98}.
However,  smaller sample sizes and the observational difficulties due to
the more complex nuclear morphologies of early-type spirals and the decreasing 
contrast between a central star cluster and a luminous bulges make this result 
somewhat uncertain. Typical luminosities of NCs are in the range 
$10^6$ - $10^7\lsun$ (Paper~I) which makes them much brighter than average 
stellar clusters in the disks of nearby spiral galaxies \citep{lar02}, and 
comparable to young ``super star clusters'' in luminous merger pairs 
\citep[e.g.][]{whi99} or circumnuclear starforming rings in spiral galaxies 
\citep[e.g.][]{mao01}.

Despite the recent progress, the formation mechanism of 
NCs remains largely a mystery. So far, no satisfying 
explanation has been put forward to explain the high gas densities 
that must have been present to enable the in-situ formation of such compact
and luminous objects in the nuclei of these disk galaxies with shallow
potentials.
Plausible alternatives to in-situ formation include the infall of evolved
clusters, possibly combined with more recent star formation events due to 
the accretion of gas in the galaxy nucleus. In order to gather evidence for or 
against any of these scenarios, it is essential to identify the stellar
population(s) that comprise the NC, and thus to constrain
the cluster formation history. Another crucial piece of information in this
undertaking is an accurate estimate for the cluster mass.

The stellar masses of NCs can be derived - under the assumption 
of spherical symmetry - if one has accurate knowledge of (a) the 
stellar velocity dispersion and (b) the cluster light distribution. 
This kind of analysis has been successfully applied to NCs \citep{boe99} 
as well as young clusters in the disks of nearby galaxies \citep{smi01,men02}.
Our team has obtained high-resolution spectroscopy of $\approx 15$ clusters
from the sample of Paper~I. These data will address point (a) above, i.e. 
to accurately measure the stellar velocity dispersion of the clusters 
\citep[][Walcher et al. 2004, in prep.]{wal03}. The purpose of the present
paper  is to discuss point (b), namely the size and spatial structure of the
NC sample of Paper~I. Because these objects are only barely resolved even
with HST, one has to carefully correct for the effects of the instrumental
point-spread function (PSF). Because standard deconvolution algorithms are
not well suited in the realm of ``almost point sources'', we have used
the specialized software package \ish\ \citep{lar99} to find the
best match between the data and PSF-convolved analytical models for
the cluster shapes. We describe the details of this exercise in 
\S~\ref{sec:sizes}, and compare the resulting structural parameters of NCs 
to those of other cluster populations in \S~\ref{sec:compare}.

Because nuclear star clusters are a relatively new aspect of
galaxy morphology, it seems a logical approach to search for correlations 
between the properties of NCs and those of their host galaxies. 
As has been shown by \cite{car99} for spirals of earlier Hubble type, such 
empirical correlations might reveal clues about the mechanism(s)
that regulate the gas supply in the very centers of spiral galaxies,
and will certainly improve our understanding of nuclear activity
in spirals. Moreover, because the presence of a NC appears to be a common 
feature in all types of spiral galaxies, the dynamical processes that 
lead to nuclear starbursts will be an important contribution to the 
discussion on bulge formation scenarios and a possible evolutionary 
connection between the various Hubble types. 
For these reasons, we present in \S~\ref{sec:corrs} a number
of correlation plots between NC sizes/luminosities
and various host galaxy properties such as size, Hubble type, HI mass,
star formation rate, or total magnitude. We summarize our 
results in \S~\ref{sec:summary}.

%%%%%%%%%%%%%%%%%%%%%%%%%%%%%%%%%%%%%%%%%%%%%%%%%%%%%%%%%%%%%%%%%%%%%%
%%%%%%%%%%%%%%%%%%%%%%%%%%%%%%%%%%%%%%%%%%%%%%%%%%%%%%%%%%%%%%%%%%%%%%
\section{Structural Analysis of Nuclear Star Clusters} \label{sec:sizes}
%%%%%%%%%%%%%%%%%%%%%%%%%%%%%%%%%%%%%%%%%%%%%%%%%%%%%%%%%%%%%%%%%%%%%%
%%%%%%%%%%%%%%%%%%%%%%%%%%%%%%%%%%%%%%%%%%%%%%%%%%%%%%%%%%%%%%%%%%%%%%
\subsection{Methodology} \label{subsec:method}
%%%%%%%%%%%%%%%%%%%%%%%%%%%%%%%%%%%%%%%%%%%%%%%%%%%%%%%%%%%%%%%%%%%%%%
Star clusters in all but the closest galaxies are marginally resolved
sources, even with the excellent resolution of HST images. In order to
measure reliably their intrinsic sizes and structural parameters, it
is therefore crucial to have both observations with high
signal-to-noise ratio (SNR) {\it and} accurate knowledge of the
instrumental point-spread function (PSF). Insufficient sampling of the
PSF can complicate matters further. For example, the pixel size of the
WFPC2 ($0.1\as$ and $0.0455\as$ for the WF and PC chips, respectively)
in general does not provide Nyquist sampling of the HST resolution
element $R = \lambda /D$. While sub-pixel dithering can alleviate this
problem to some extent, HST observations of extragalactic star
clusters in all but the closest galaxies are almost always confined 
to the regime of ``marginally resolved''.

In this situation, none of the conventional deconvolution methods such
as the Maximum Entropy Principle \citep*{bur83} or the
Richardson-Lucy algorithm \citep{ric72,luc74} work reliably. Instead,
differential aperture photometry \citep[e.g. the magnitude difference $m_2 -
m_4$ between concentric apertures of 2 and 4 pixels, ][]{hol92} of
stellar clusters and comparison to individual stars broadened by
Gaussians has been used to estimate cluster sizes.  However, a
Gaussian is generally a poor approximation to the surface brightness
profile of stellar clusters. In particular, the wings of a Gaussian
fall off too rapidly, and cannot adequately describe the outer regions
of either old globular clusters (GCs) in the Milky Way or young GCs
in external galaxies.

More recently, a number of authors \citep[e.g.][]{kun98,lar99,ch01}
have independently developed techniques to determine more reliably the
structure of stellar clusters in external galaxies.  The various
methods are conceptually similar in that they assume a parametric
model for the intrinsic light distribution. A particular model is
considered to be a good description of reality if - after convolution
with the instrumental PSF - it matches the observations. The ``best''
model is the one that minimizes the $\chi^2$-difference to the data.
The most widely used analytic models for the projected density profile
of stellar clusters are those of \cite{kin62}:
\beq
f(z) =\left\{
\begin{array}{ll}
( \frac{1}{\sqrt{1+z^2} } - \frac{1}{\sqrt{1+c^2} })^2 & {\rm for} \> z < c \\
0 & {\rm for} \> z \geq c\ \ . \\
\end{array}
\right.
\eeq
Here, $z$ is defined by $z^2 = a_1 x^2 + a_2 y^2 + a_3 xy$, with $x$ and 
$y$ denoting the coordinates relative to the center of the profile and 
constants $a_1$, $a_2$, and $a_3$ which depend on the major axis, 
ellipticity and orientation of the model. The so-called concentration index 
$c$ is the ratio between the tidal radius $r_{\rm t}$ and the core radius 
$r_{\rm c}$.

In the regime of marginal resolution, the results of \cite{ch01} have 
shown that unless the SNR of the
observations is exceptionally high ($\gta 500$), it is very difficult
to distinguish between King models of different concentration indices.
On the other hand, they find that the effective radius \re , i.e. the
radius which contains half the cluster light in projection, can be fairly
robustly
estimated.  The value of \re\ does not depend strongly on the exact
shape of the assumed model. This is in agreement with the results of
\cite{kun98} and has been extensively tested by \cite{lar99}. In what
follows, we will describe our attempts to measure the effective radii
of the NCs found in the survey of Paper~I.

For our analysis, we have made use of S\"oren Larsen's software package
\ish , which is described in detail in \cite{lar99}.  This software
package has been well tested in a number of studies
\citep[e.g.][]{men02,lar02c},
and has proven to be a robust and user-friendly tool for structural
studies of compact star clusters.  In brief, the program convolves a
two-dimensional analytical model (e.g. Gauss-, Hubble-, Moffat-, or
King-profiles) with a user-provided PSF, and finds the particular
parameter set that best describes the observations, including the
exact location of the cluster centroid on sub-pixel scales. Comparison
with the pure PSF (i.e., a PSF-convolved $\delta$-function) allows
to assess whether the clusters are resolved at HST resolution.
Because \re\ is not defined for functions with
divergent integrals such as Hubble-type profiles, we have only
attempted to fit the data with the above-mentioned King models as 
well as Moffat profiles which are defined as
\beq
f(z) = \frac{1}{(1+z^2)^i}\ \ ,
\eeq
with $z$ defined as for the King model, and $i>1$ denoting the so-called
power index.

In order to construct accurate PSFs for our analysis, we have used the
\ttim\ software \citep{kri01}, which accounts for the spectral
passband used, the plate scale variations across the field of view,
and the spectral energy distribution (SED) of the source. Although all
of our images were taken through the same filter (F814W), we have
constructed a separate PSF, according to the exact position of the
NC on the detector array. We note, however, that this
only marginally affects the results because the WFPC2 PSF varies only
slightly across the PC field. The choice of the correct SED for
constructing the PSF is somewhat more critical in estimating the
NC sizes. Unfortunately, spectroscopic information is
available for only a limited number of NCs. Early results
of our spectroscopic follow-up survey \citep{wal03} show that the
light of many NCs is dominated by a stellar population 
in the age range 50-$300\myr$. 
This agrees with a number of case studies of individual clusters, e.g.
in M\,33 \citep{gor99}, NGC\,2403 \citep{dav02}, IC342 \citep{boe99},
or NGC\,4449 \citep{boe01}. We therefore chose the SED of an A5 star
for constructing the \ttim\ PSF for all objects. We have verified that
an alternate choice for the cluster SED, e.g. that of a cool giant
with spectral type M3, does not significantly change the \ish\
results.

%%%%%%%%%%%%%%%%%%%%%%%%%%%%%%%%%%%%%%%%%%%%%%%%%%%%%%%%%%%%%%%%%%%%%%
\subsection{Results} \label{subsec:fitresults}
\begin{figure}[htb]
\vspace*{1cm}
\centerline{\epsfig{file=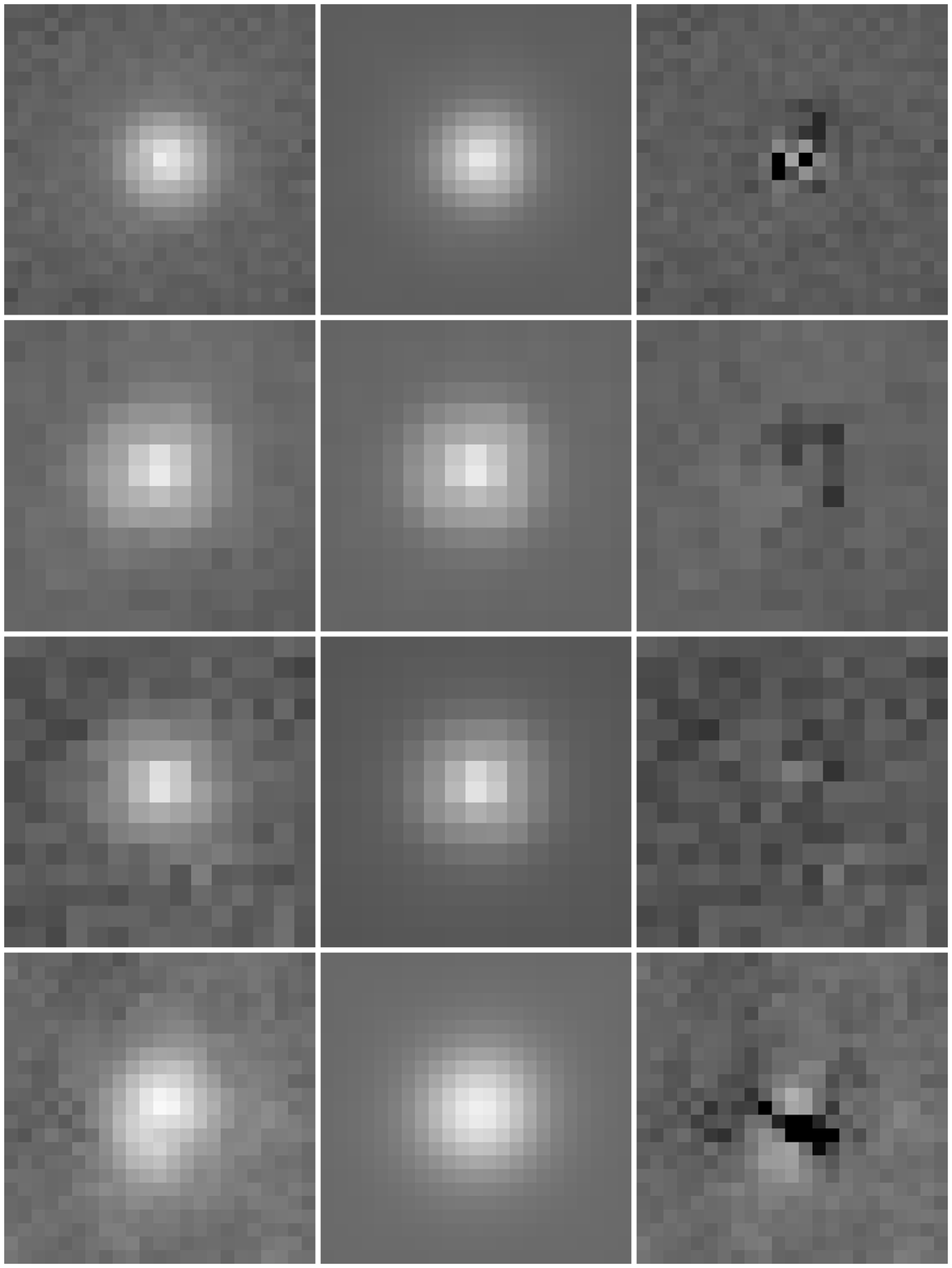, width=12cm} }
\vspace*{3cm}
\caption[f1.eps]{\label{fig:ishape_mosaic} 
       Some typical examples of \ish\ fits.
       The three panels of each row show (from left to right) the data,
       the best \ish\ fit (analytic model convolved with \ttim\ PSF), and
       the residuals (data $-$ fit). The field size of each panel corresponds 
       to the aperture radius $R_u$ (Paper~I) over which the NC 
       dominates the emission. From top to bottom, the objects are NGC\,2805,
       NGC\,3445, NGC\,3906, and NGC\,4618. The last was excluded from
       the analysis because of its complex morphology which prevented a
       good fit.}
\end{figure}

%%%%%%%%%%%%%%%%%%%%%%%%%%%%%%%%%%%%%%%%%%%%%%%%%%%%%%%%%%%%%%%%%%%%%%
We have used \ish\ to fit all 59 NCs of Paper~I within the aperture radius 
that is dominated by emission from the NC (at least 5 \re ). 
Some typical examples that illustrate the \ish\ results are presented 
in Figure~\ref{fig:ishape_mosaic}. We used only circularly symmetric
models within \ish\ because visual inspection of the fit results
confirmed that the majority of clusters show little evidence for
flattening. In a few cases, the clusters are asymmetric or have a
complex morphology, most likely due to dust lanes or faint companion
clusters. These objects were excluded from the \ish\ analysis. In
addition, we excluded those clusters with insufficient SNR to obtain 
reliable fits, i.e. with SNR $<$ 40 \citep{lar99}. 

The remaining 39 clusters were fit using King profiles with
$c\equiv r_{\rm t}/r_{\rm c}$ values of 15, 30, and 100, as well as Moffat
profiles with power indices of $i=1.5$ and $i=2.5$.
In principle, the galaxy light underlying the NC must be subtracted
before fitting the cluster profile. The \ish\ code simply subtracts 
a constant background level, measured at the egde of the fitting aperture.
While this algorithm certainly is less than perfect, and more sophisticated
methods to fit the surface brightness profile of the galaxy disk could
be employed, we have verified that the resulting values for \re\ are not
systematically biased by this shortcoming. In fact, the cluster magnitudes
calculated by integrating the best-fit \ish\ model agree well with those 
listed in Paper~I which were derived with a more careful estimation of 
the galaxy background (see \S~\ref{subsec:mwgc_comp}). The median (mean) 
difference between these two sets of numbers is 0.07 (0.11) magnitudes when
integrating the models over the same aperture used for the Paper~I
photometry.

The best-fitting models
and the resulting effective radii, both in angular and physical dimensions, 
are summarized in Table~\ref{tbl:props}. The majority (29/39) of NCs 
have \re\ $< 5\pc$, with median (mean) values of $3.5\pc$ ($5.1\pc$). 
As discussed further in \S~\ref{sec:compare}, these values are comparable to 
the typical size of old globular clusters in the Milky Way and other nearby 
galaxies. We emphasize that all 
clusters are resolved, i.e. models that simply rescale and position the 
\ttim\ PSFs are inconsistent with the data by a large margin. For
completeness,
we also list in Table~\ref{tbl:props} the best-fitting analytic model. 
We caution, however, that in some cases, other models may also 
be consistent with the data. However, this has little impact on the
uncertainties of the \re\ measurements, as explained further below. Our goal 
is not to determine the detailed shape of the cluster profile, but merely 
to obtain a robust estimate for its size.

The uncertainties listed in Column~7 of Table~\ref{tbl:props} represent
the range of \re\ for the best-fitting model profile that yields 
acceptable fits to the data. Here, ``acceptable'' is defined by the 
$3\sigma$ confidence limit for one free parameter, 
i.e. $\Delta\chi^2 < 9$ \citep[e.g.][]{pre92}.
Our analysis is conservative in the following sense: before evaluating
the quality of the model fits, we rescaled all $\chi^2$ values such that 
the best \ish\ model for a given galaxy had a $\chi^2$ value equal to the
number of degrees of freedom, i.e. the number of data points (pixels in
the aperture) minus the number of free parameters. This rescaling was necessary
because \ish\ appears to underestimate the intrinsic errors in the cluster 
images. As a result, many of the best $\chi^2$ values do not formally
represent ``good'' models according to $\chi^2$-statistics. As
the scale factor was always smaller than one, the $\chi^2$ rescaling is 
equivalent to increasing the noise level of each data point or pixel, and
consequently broadens the range of models that appear consistent with
the data. Unfortunately, this rescaling of errors made a more rigorous 
statistical comparison between the different analytic profiles impossible.

\begin{figure}[htb]
\centerline{\epsfig{file=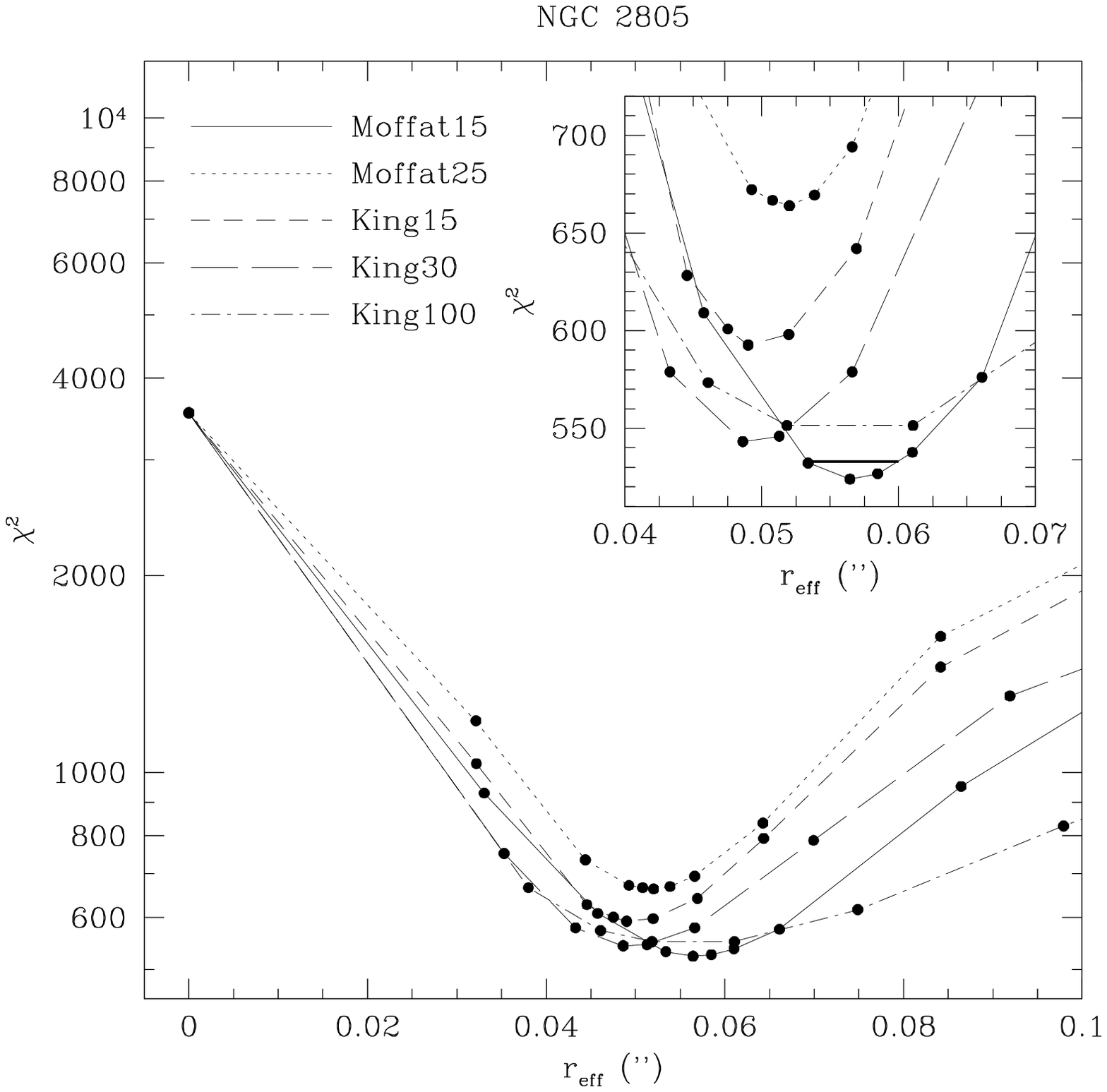, width=15cm} }
\caption[f2.eps]{\label{fig:2805_allmodels}
	\ish\ fit results for NGC\,2805. Shown are the $\chi^2$ values of each
	fit as a function of effective radius for all analytic models explored.
	These values have been rescaled such that the minimum $\chi^2$
	matches the number of degrees of freedom (see text).
	The inlay panel shows that only the MOFFAT15 model with \re\ values
	of $0.056\as \pm 0.006\as$ falls below the $3\sigma$ confidence
	limit for 5 free parameters ($\Delta\chi^2=18.2$, indicated by the dashed
	horizontal line). Within the MOFFAT15 model, the range of
	acceptable vaues for \re\ is marked by the solid horizontal line
	which denotes the $3\sigma$ confidence limit for a single free parameter
	($\Delta\chi^2=9$).}
\end{figure}

To illustrate the analysis, Fig.~\ref{fig:2805_allmodels} shows 
for the example of NGC\,2805 the rescaled $\chi^2$ values as a function
of \re\ for all models considered. In this case, a Moffat profile
with a power index of 1.5 (MOFFAT15) and effective radius 
\re\ $= 0.056\as \pm 0.003\as$ yields the best fit to the data.
The $3\sigma$ confidence limit and the corresponding range of \re\
is indicated by the horizontal line in the inlay panel. 
As mentioned before, the effective radius is a rather robust quantity, i.e.
different models yield similar values for \re , even if they are not
formally consistent with the data according to the above criterion for
$\Delta\chi^2$. In the case of NGC\,2805, this can be seen in 
Fig.~\ref{fig:2805_allmodels}: the ``best''
values for \re\ of all five models are within 20\% of each other which is
also the typical uncertainty listed in Table~\ref{tbl:props}.  

%%%%%%%%%%%%%%%%%%%%%%%%%%%%%%%%%%%%%%%%%%%%%%%%%%%%%%%%%%%%%%%%%%%%%%
\section{Discussion of Structural Parameters} \label{sec:compare}
%%%%%%%%%%%%%%%%%%%%%%%%%%%%%%%%%%%%%%%%%%%%%%%%%%%%%%%%%%%%%%%%%%%%%%
%
%%%%%%%%%%%%%%%%%%%%%%%%%%%%%%%%%%%%%%%%%%%%%%%%%%%%%%%%%%%%%%%%%%%%%%
\subsection{Comparison to Milky Way Globular Clusters}\label{subsec:mwgc_comp}
%%%%%%%%%%%%%%%%%%%%%%%%%%%%%%%%%%%%%%%%%%%%%%%%%%%%%%%%%%%%%%%%%%%%%%
A natural question to address with the data we have in hand is that of
a possible size-luminosity relation for NCs. Since NCs in late-type disks 
are potential candidates for the seeds of bulge formation,
they might follow the mass-radius correlation established
for more massive ellipsoidal stellar populations, i.e. elliptical
galaxies and bulges of intermediate- and early-type spirals 
\citep[e.g.][]{kor85,ben92}. On the other hand, the
old GCs in the Milky Way -- which constitute the best-studied
family of stellar clusters -- are remarkably free of any such trend: the
effective radii of Galactic GCs are independent of cluster
luminosity or mass \citep{van91,djo94,mcl00}. The same is true for
GCs in the nearby galaxies M\,31 \citep{bar02}, M\,33
\citep{lar02b}, and NGC\,5128 \citep{har02}.

\begin{figure}[htb]
\centerline{\epsfig{file=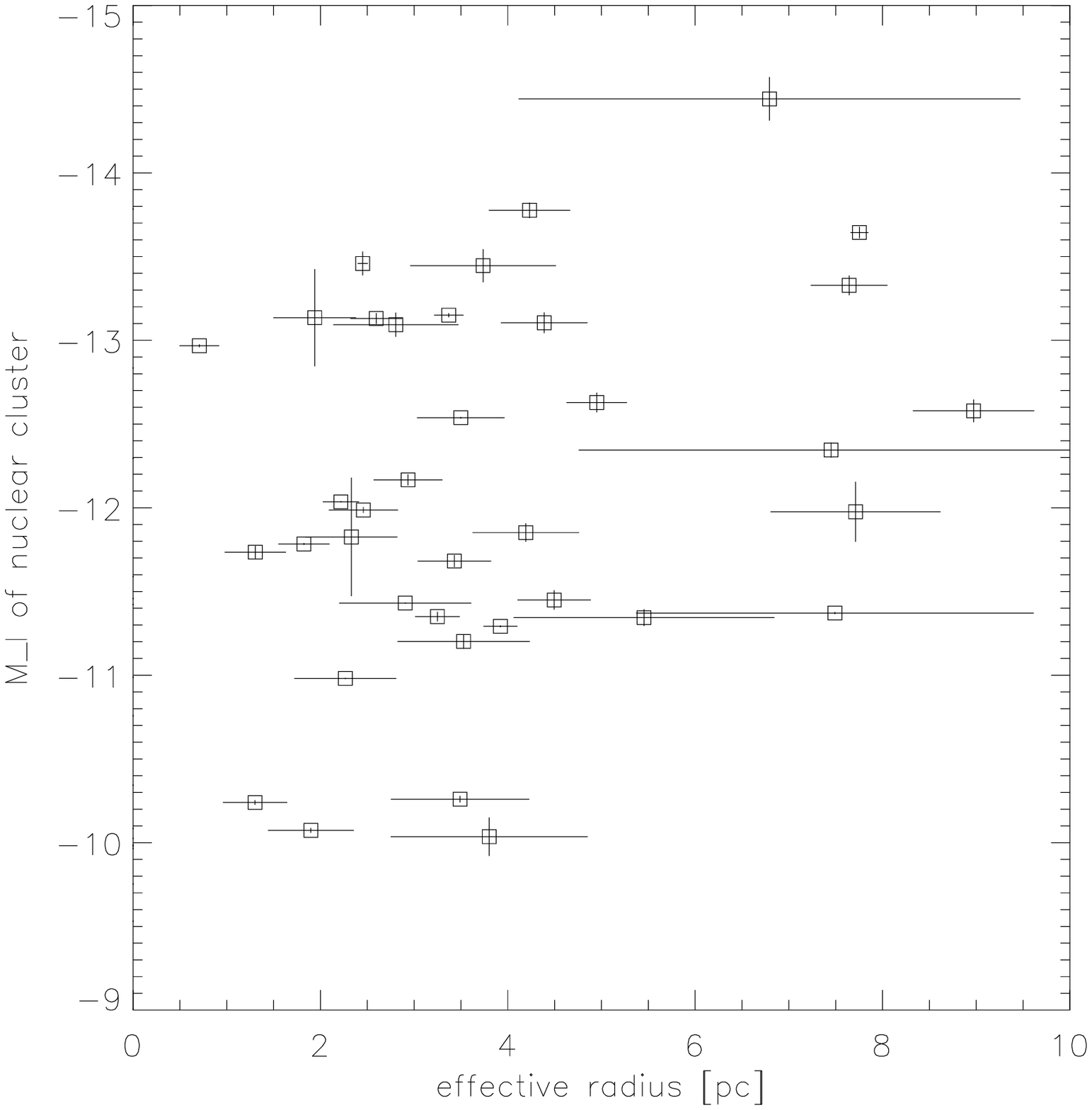, width=9cm} }
\caption[f3.eps]{\label{fig:brightbig}
	Comparison of NC effective radius and absolute magnitude.
	The value of \re\ is not a strong function of cluster luminosity:
	clusters of similar size span a wide range in magnitude. Two clusters
	with large \re\ values have been omitted from the plot in order to
	avoid crowding of the data points, they are the NCs in NGC\,450 
	(\re$=24.3\pc$, \mic$=-11.96$) and NGC\,4625 (\re$=28.8\pc$, \mic$=-10.63$).}
\end{figure}

In Fig.~\ref{fig:brightbig}, we plot the absolute I-band magnitude of the
NCs, \mic , against their effective radius, \re .
The procedure for
measuring \mic\ is described in detail in Paper~I. In brief, the challenge
is to determine the light contribution of the stellar disk underlying the
NCs. For this, we use two limiting models for the inward 
extrapolation of the galaxy disk profile. The \mic\ values in Column~5
of \ref{tbl:props} represent the average of the resulting two numbers; half
their difference determines the size of the error bars in
Fig.~\ref{fig:brightbig}.
Aside from a handful of bright clusters with effective radii larger than 
average, the bulk of the sample---28 of 39 clusters---is found in the narrow
range  $1\pc \le r_{\rm e} \le 5\pc$, 
while the clusters range over a factor of $\sim 100$ in luminosity, from 
$-9 \gta$ \mi\ $\gta -14$. As we discuss below, relating luminosity to mass 
is nontrivial for NCs, but taken at face value, the situation 
illustrated in Fig.~\ref{fig:brightbig} is more reminiscent of GCs 
than of the bulges in early-type spirals.

It is not only the absence of a correlation between size and luminosity that
appears to connect the NCs in our sample to GCs. 
As noted in \S~\ref{subsec:fitresults}, the distribution of \re\ for 
our NC sample is quantitatively similar
to that of Galactic GCs. This is shown explicitly
in Fig.~\ref{fig:hists}, where we also compare the distributions of
I-band magnitude for our NC sample and for 111 Milky Way GCs
without an obvious or suspected core-collapse morphology, i.e.,
with well-determined \re\ values.\footnote{The Milky Way GC data are taken 
from the February 2003 version of the online catalogue compiled by
\cite{har96},
which contains 141 clusters with known values for angular half-light
radius and heliocentric distance, King concentration index, absolute V-band
magnitude, and $(B-V)$ color and excess. For 96 of these, Harris also
gives an apparent $(V-I)$ color, which we correct according to
$E(V-I)=1.25\,E(B-V)$ \citep{car89} in order to derive $M_I$ from $M_V$.
We simply assign the average $(V-I)_0=0.95$ to the other 45 clusters.}
Evidently, although the median $M_I$ of NCs is fully 3--4 mags brighter 
than  that of the Milky Way GCs, the widths of the two luminosity distributions
(which corresponds to a distribution in mass for the GCs) are comparable. 
The median effective radius of the NCs is
$\overline{r_e}=3.5\ {\rm pc}$ (interquartile range 2.4--5.0 pc) and that of
the GCs is $\overline{r_e}=3.3\ {\rm pc}$ (interquartile range 2.4--5.2 pc).
A two-sample Kolmogorov-Smirnov test confirms statistically the visual
impression that the two \re\ distributions are identical, with a probability
$P_{\rm KS}=0.75$ that they have been drawn from the same parent distribution.
We further note that the GCs in M31, M33, and NGC 5128 have essentially the
same median \re\ as the clusters in Fig.~\ref{fig:hists}, and the
same independence from cluster luminosity (see the references above).

Figure~\ref{fig:fp} provides an alternate illustration of this basic
point, in
a representation that allows for direct comparison with larger ellipsoidal
stellar systems and comes closer to the ``fundamental plane'' typically
plotted for such systems. Here, we show the average surface brightness,
$I_{\rm e}\equiv L_{\rm I}/(2\pi r_{\rm e}^2)$, vs. effective radius \re\ for
the NCs, the 111 Galactic GCs from Harris' (1996) catalogue,
and a number of E galaxies, spiral bulges, and dwarf Ellipticals (dE's)
from the compilation of \cite{bur97}. The position of Milky Way GCs and
NCs to the left of the plot shows again that the latter
have essentially the same size as the former, but are substantially
brighter and therefore at higher surface brightness.

Of additional interest is the disconnect between the GCs and
NCs on the one hand and the elliptical galaxies and spiral bulges on
the other, a dichotomy first noted by \cite{kor85}. As we mentioned at 
the beginning of this discussion, the bulges and galaxies show a 
size-luminosity relation, while GCs do not.
%{\bf [Note: how about plotting the bulges with a point type
%different from the E galaxies in this Figure? They may be offset.]}.
Equivalently, GCs have an enormous spread in surface brightness,
while bulges and ellipticals at a given \re\ appear much more uniform.
Although this fact is still not well understood, the alignment of our
NCs with the GCs in Figure \ref{fig:fp} would seem to underline a closer
link between NCs and GCs, as
opposed to one between NCs and classical galaxy bulges.

For completeness, we have included in Figure~\ref{fig:fp} data for
the nuclei of 5 nucleated dE galaxies in Virgo, from the sample of
\cite{geh02}. As those authors have already demonstrated, with a different
version of the ``fundamental plane'', the dwarf-elliptical  nuclei are also
more similar to star clusters than to galaxy bulges.

%%%%%%%%%%%%%%%%%%%%%%%%%%%%%%%%%%%%%%%%%%%%%%%%%%%%%%%%%%%%%%%%%%%%%%
%%%%%%%%%%%%%%%%%%%%%%%%%%%%%%%%%%%%%%%%%%%%%%%%%%%%%%%%%%%%%%%%%%%%%%
\subsection{Possible Interpretations}
\label{subsec:brightbig}
\begin{figure}[htb]
\centerline{\epsfig{file=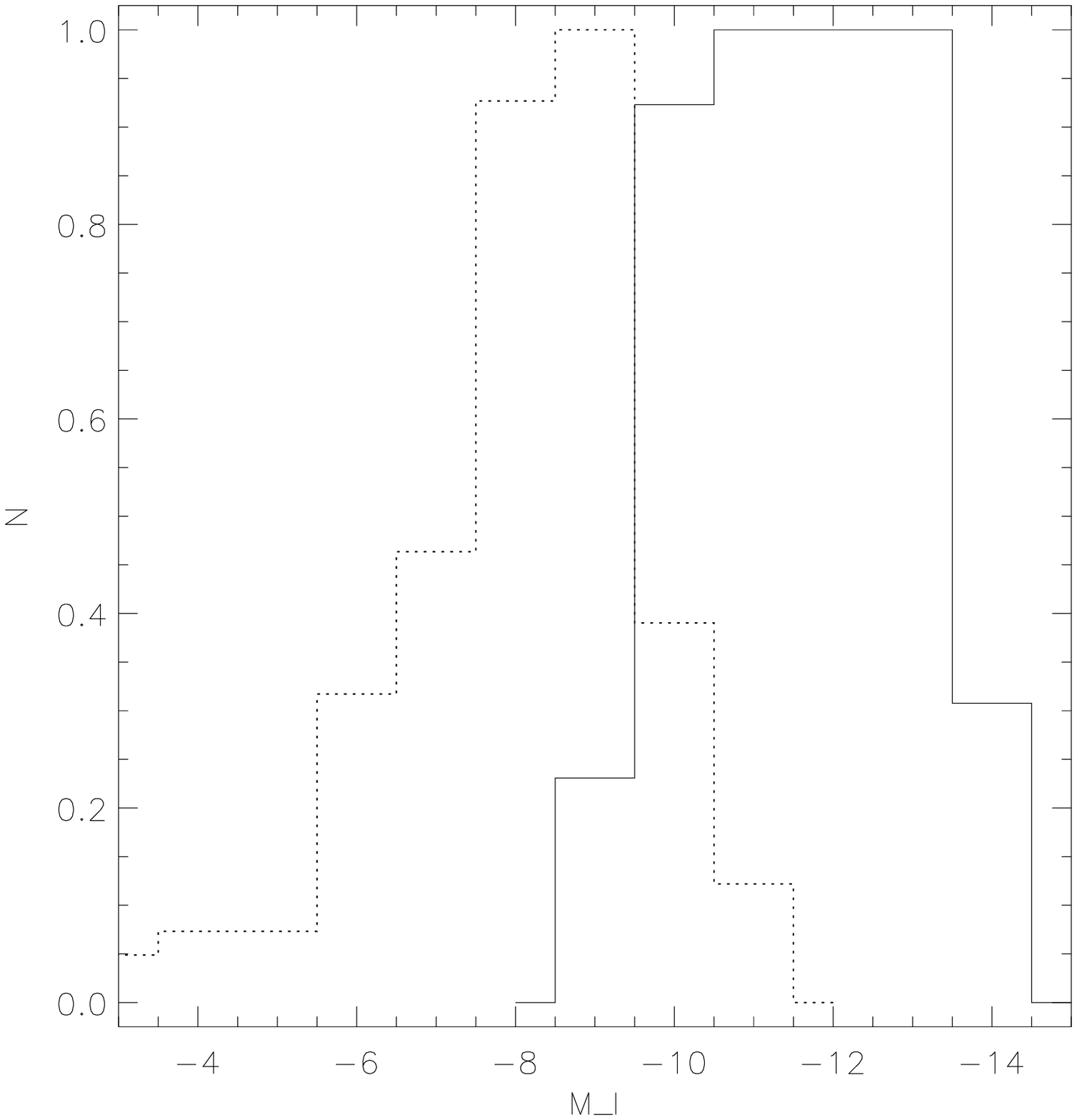, width=9cm}
	    \epsfig{file=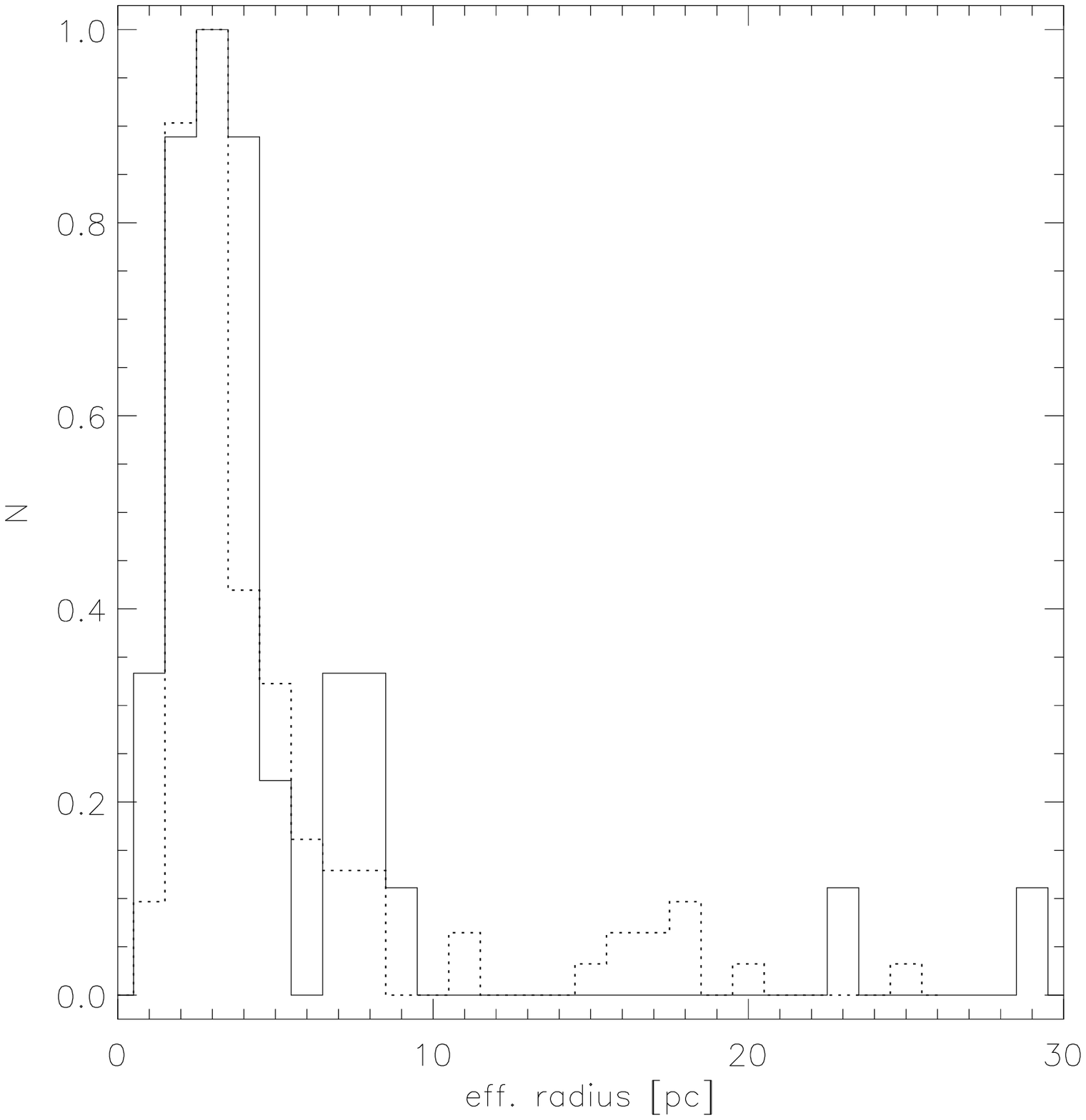, width=9cm} }
\caption[f4.eps]{\label{fig:hists}
	 Left: distribution of total absolute I-band magnitude for all
	 NCs of Paper~I (solid line), compared to that of the Milky Way
	 GCs (dotted line).
	 Right: distribution of effective radii for all NCs with
	 good \ish\ fits (solid line), compared to the Milky Way GCs
	 (dotted line). All histograms have been normalized to a
	 peak value of 1.}
\end{figure}

\begin{figure}[htb]
\centerline{\epsfig{file=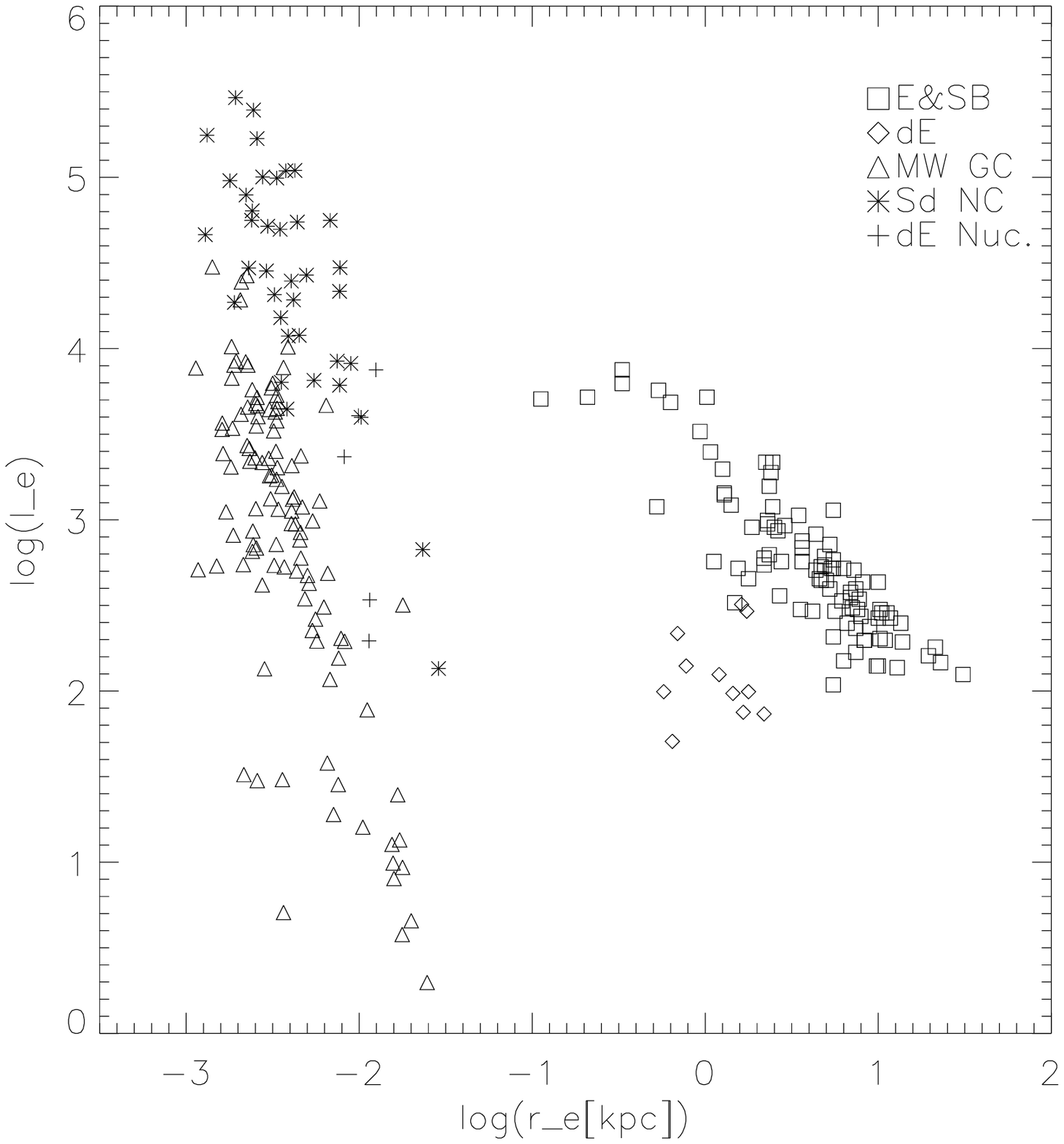, width=12cm} }
\caption[f5.eps]{\label{fig:fp}
	 Comparison of NCs with
	 other spheroidal populations in the \re\ vs. $I_{\rm e}$ plane. The
	 data are from the following sources: Galactic GCs:
	 \cite{har96}, Ellipticals, spiral bulges, and dE galaxies:
	 \cite{bur97}, dE nuclei: \cite*{geh02}. All photometry has been
	 converted to I-band, assuming the following colors: B$-$I=2.0
	 for E, S0, and dE galaxies from the \cite{bur97} database,
	 V$-$I=0.8 for the dE nuclei of \cite{geh02}, and V$-$I=0.95 for those
	 GCs which have no value for \mi\ listed in the \cite{har96} catalog
	 (see \S~\ref{subsec:mwgc_comp}). }
\end{figure}

%%%%%%%%%%%%%%%%%%%%%%%%%%%%%%%%%%%%%%%%%%%%%%%%%%%%%%%%%%%%%%%%%%%%%%
The apparent similarity between NCs and GCs could be an important clue to the
nature and origin of the NCs themselves. First, however, their identification
as an extension of the GC sequence to brighter magnitudes
(Fig.~\ref{fig:hists})
and higher surface brightness (Fig.~\ref{fig:fp}) needs to be understood in 
terms of {\it mass} rather than luminosity alone.
Of particular concern in this context is the interpretation of the spread of NC
absolute magnitudes in Fig.~\ref{fig:brightbig} and Fig.~\ref{fig:hists}.
We consider two extreme scenarios---neither of which we mean necessarily
to advocate as plausible {\it per se}, but which may serve to delimit further
discussion:

(1) If the stellar populations of all NCs shared a common age,
then the five-magnitude spread in \mic\ would translate directly to
a range of a factor of $\sim$100 in NC mass, intriguingly similar to the
corresponding variation among old GCs (Fig.~\ref{fig:hists}). Were
Fig.~\ref{fig:fp} then to be re-plotted
using mass surface densities, the NC locus would simply
shift vertically, by an amount depending on the typical NC age
(through the $I$-band mass-to-light ratio). This would strengthen the
connection of NCs to GCs and sharpen their distinction from galaxy bulges.

(2) If instead all NCs had roughly the same stellar mass, then the spread
in \mic\ would have to arise from a range in luminosity-weighted age,
with fainter cluster magnitudes (corresponding to systematically older
ages) being purely the result of stronger population fading. If
this were true, then Fig.~\ref{fig:fp} re-drawn in terms of mass surface
densities would have the NC locus contracted into a relatively narrow line
of slope $-2$. The small variation in NC effective radius would then be
largely the result of a small mass range in our sample, and it would reveal
little about the (non)existence of a mass-radius relation among these
clusters.

\begin{figure}[htb]
\centerline{\epsfig{file=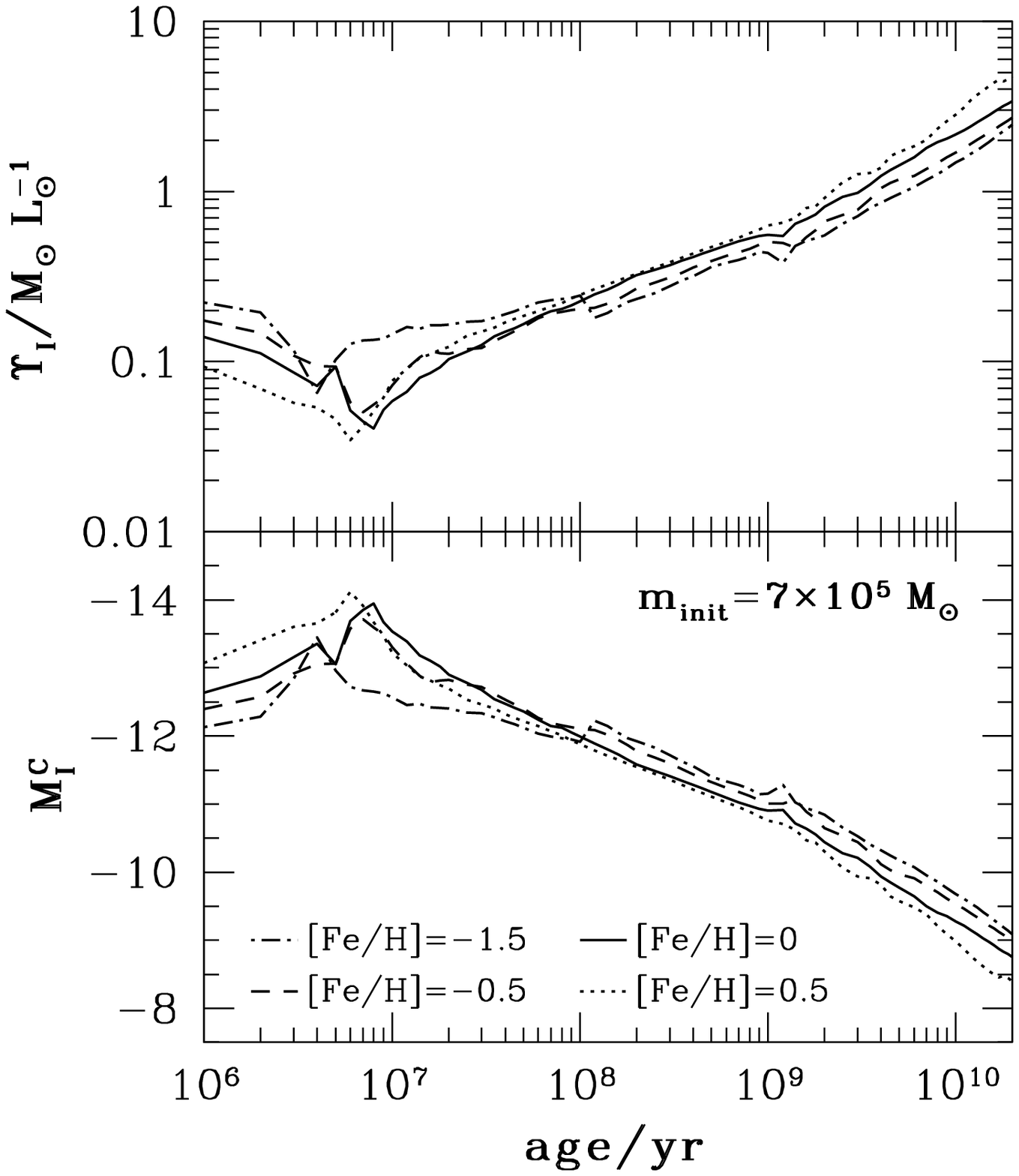, width=16cm} }
\caption[f6.eps]{\label{fig:pegase}
	 $I$-band mass-to-light ratio (top) and absolute $I$-band magnitude
	 for a stellar cluster of initial mass $7\times10^5\ M_\odot$,
	 formed in a single burst of star formation with the IMF of
	 \cite{kro93} between lower- and upper-mass cut-offs of
	 $0.1\ M_\odot$ and $120\ M_\odot$, as derived from the PEGASE
	 code of \cite{fio97}. The predictions include the effects of
	 mass loss due to stellar winds and supernovae, but not the
	 effects of dunamycal evolution. The four lines of each panel
	 correspond to different metallicities as indicated. }
\end{figure}

In order to gain some quantitative feel for the viability of these two
alternatives, we have used the PEGASE population-synthesis code
\citep{fio97} to calculate the evolution of $I$-band
mass-to-light ratio and absolute magnitude for a stellar cluster of mass
$7\times10^5\ M_\odot$, formed in a single burst of star formation with the
IMF of \cite{kro93} between lower- and upper-mass cut-offs of  $0.1\ M_\odot$
and $120\ M_\odot$. The results, for a range of assumed metallicities, are
presented in Fig.~\ref{fig:pegase}. Note that the mass-to-light ratio in 
the top panel is independent of the assumed total mass of the cluster, while
the $I$-band magnitude in the bottom panel is of course sensitive to this
parameter: for a given metallicity, the absolute magnitude of a single-burst
stellar population of any mass $m$ is obtained, as a function of age, by
adding $-2.5\,\log(m/7\times10^5\,M_\odot)$ to the appropriate curve
in Fig.~\ref{fig:pegase}. The calculations for $\Upsilon_I$ and \mic\ both take
into account the mass lost from a cluster due to stellar evolution, i.e.,
winds and supernova explosions.

If the NCs were in fact a roughly coeval population of clusters with 
a narrow range in mass-to-light ratio $\Upsilon_I \equiv M/L_I$ 
(scenario (1) above), then they could in principle have masses similar to 
the Galactic GCs if $\Upsilon_I$ were some 15 times smaller for NCs than 
for GCs (to explain the 3-magnitude difference in the peaks of the \mic\ 
distributions of Fig.~\ref{fig:hists}). The top panel of Fig.~\ref{fig:pegase} 
shows that a ``typical'' GC, with an age of 13 Gyr and ${\rm [Fe/H]}=-1.5$, has
$\Upsilon_I\simeq 1.75\ M_\odot\,L_\odot^{-1}$ in these
models.\footnote{The observed number is about
$1.2\ M_\odot\,L_\odot^{-1}$, which follows from an average $V$-band value of
1.45 in solar units \citep{mcl00}, and a typical $V-I$ color of 0.95
(see \S~\ref{subsec:mwgc_comp}).}
The same graph then shows that the NCs would have to be of order $\sim10^7$
years old (to within a factor of two or so) in order to have
$\Upsilon_I\sim0.1\ M_\odot\,L_\odot^{-1}$, and thus a mass distribution 
similar to that of the GCs. Interestingly, this would make them potential 
analogues of the super star clusters forming in many nearby interacting 
and starburst galaxies \citep[e.g.,][]{whi99}. In this context, it is
worth noting that these super star clusters tend to have effective radii
$r_e\sim2-10$ pc that do not appear to depend strongly on
luminosity---reminiscent of NCs
and GCs both \citep[e.g.,][]{barth95, zepf99}. Nevertheless, it is difficult
to imagine how such a large fraction of late-type disks could have conspired
simultaneously to form nuclear clusters such a short time ago.

If, on the other hand, scenario (2) were correct, and the spread in NC
luminosity reflected an age spread among single-burst stellar populations
with essentially a single mass, then the bottom panel of Fig.~\ref{fig:pegase}
suggests that this common mass would have to be near $7\times10^5\ M_\odot$,
and the age range exceedingly broad: from $10^7$--$10^{10}$ years to just
barely reproduce the observed $-14\la M_I^C\la -9$ (Fig.~\ref{fig:hists}).
(A change from the ``fiducial'' mass in Fig.~\ref{fig:pegase} by a factor of
even 2.5 in either direction would shift all the \mic\ curves up or down by
a full magnitude, leaving an incomplete overlap with the observed NC
magnitude distribution.)

Even as oversimplified as these two scenarios are, it is not possible to
discriminate further between them without additional information. We have
therefore obtained short-wavelength spectra for a number of the NCs in the
current sample, which we will fit with population-synthesis models to
constrain the clusters' ages and star-formation histories, and from which we
will extract stellar velocity dispersions to constrain dynamical mass-to-light
ratios. Although this analysis is still in progress, the first results
\citep{wal03} suggest a picture that is, not surprisingly, rather more
complicated than either of the extremes just discussed.

\cite{wal03} show that the light-dominating population of stars in 10
relatively bright NCs range in age from $\sim10^7$--$10^9$ years old. That
is, there is in fact a spread in (luminosity-weighted) age among the NCs ---
although it does not appear sufficient to explain the full range of \mic\
given the  single-burst models in Fig.~\ref{fig:pegase}, unless there is
{\it also} a significant (order-of-magnitude) spread in the cluster masses.
More importantly, however, \cite{wal03} show that the single-burst hypothesis 
itself is almost certainly incorrect: the spectra of their sample of clusters 
show some direct evidence for a mix of stellar ages, and the dynamical
mass-to-light ratios they derive are consistently higher than the value of
$\Upsilon_I$ inferred from  population-synthesis modeling, suggesting that an
underlying old (faint but massive) population of stars may be present.

In view of this, and given the similarity between GC and NC effective
radii, an obvious first-order enhancement over the single-burst idea is
the hypothesis that NC masses are in fact dominated by old stellar
populations -- perhaps {\it bona fide} GCs that have been dragged
by dynamical friction to the centers of their host galaxies -- but that
their luminosities, and hence the young ages implied by their spectra, are
due mainly to more recent star formation triggered by the infall
of a smaller mass of circumnuclear gas. A more thorough exploration of this
scenario is beyond the scope of this paper and must await the full
analysis of our spectroscopic data (Walcher et al. 2004, in preparation).
Nevertheless, it is worth noting that recent high-resolution interferometric 
studies of the nuclear gas dynamics in nearby galaxies have shown that
molecular gas flows can indeed reach to within the central few pc of the
nucleus \citep*[e.g.][]{sch03} and thus provide a possible mechanism
for the fueling of such nuclear starburst events. 
%The {\it Atacama Large Millimeter Array} (ALMA) will be needed for
%a more systematic study of the interplay between molecular gas flows and 
%nuclear star clusters.
%%%%%%%%%%%%%%%%%%%%%%%%%%%%%%%%%%%%%%%%%%%%%%%%%%%%%%%%%%%%%%%%%%%%%%
%%%%%%%%%%%%%%%%%%%%%%%%%%%%%%%%%%%%%%%%%%%%%%%%%%%%%%%%%%%%%%%%%%%%%%
\section{Correlations with Host Galaxy Properties} \label{sec:corrs}
%%%%%%%%%%%%%%%%%%%%%%%%%%%%%%%%%%%%%%%%%%%%%%%%%%%%%%%%%%%%%%%%%%%%%%
%%%%%%%%%%%%%%%%%%%%%%%%%%%%%%%%%%%%%%%%%%%%%%%%%%%%%%%%%%%%%%%%%%%%%%
%
%%%%%%%%%%%%%%%%%%%%%%%%%%%%%%%%%%%%%%%%%%%%%%%%%%%%%%%%%%%%%%%%%%%%%%
\subsection{Luminosity and Mass of the Host Galaxy}
\begin{figure}[htb]
\centerline{\epsfig{file=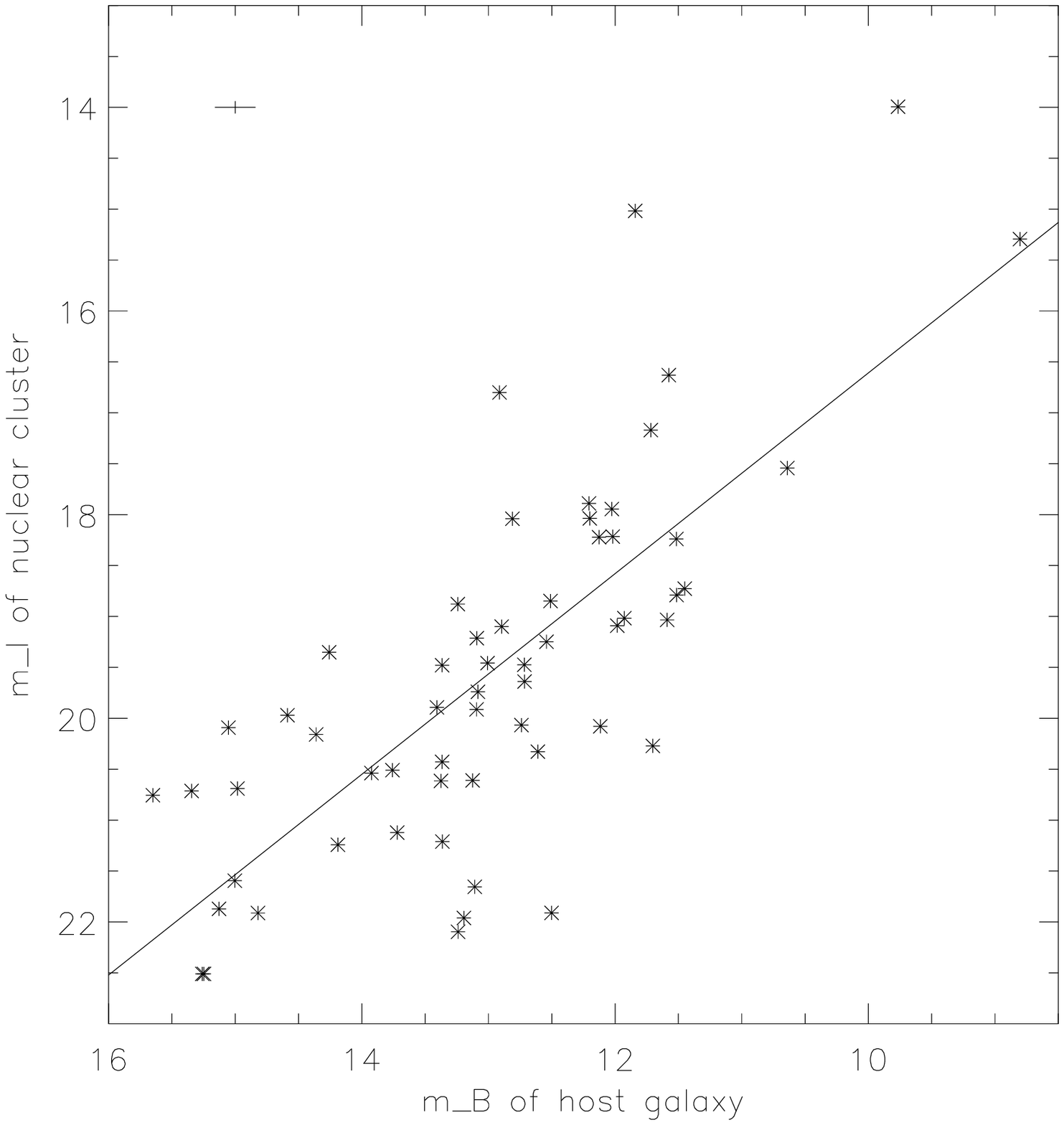, width=9cm}
	    \epsfig{file=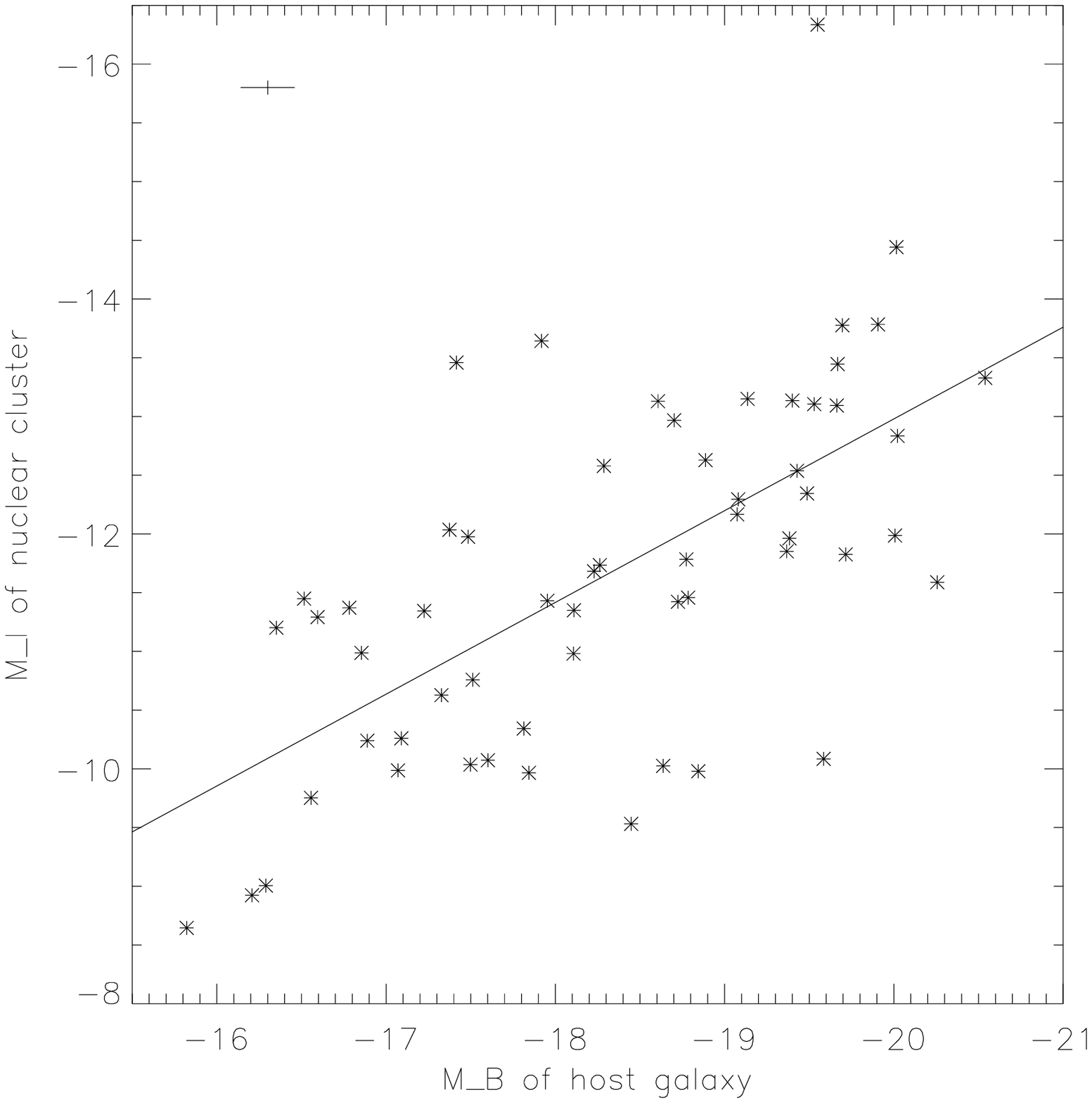, width=9cm} }
\caption[f7.eps]{\label{fig:lumcorrs}
	 Comparison between NC luminosity and
	 total luminosity of the host galaxy.
	 a) apparent I-band magnitude of NC
	 vs. apparent total B-band magnitude of the
	 host galaxy (from the LEDA database).
	 b) same as a), but for absolute magnitudes (calculated
	 using the distances listed in Paper~I). The median error
	 bar is indicated in the upper left corner of each panel,
	 excluding (for panel b) any (unknown) distance errors.
	 Both panels show a statistically significant correlation,
	 which is indicated by the the straight line in each panel.
	 The lines mark the best least-squares linear fit to the data
	 (fit coefficients and significance levels are given in
	 Table~\ref{tbl:coeff}.  }
\end{figure}

%%%%%%%%%%%%%%%%%%%%%%%%%%%%%%%%%%%%%%%%%%%%%%%%%%%%%%%%%%%%%%%%%%%%%%
The work of \cite{car98} has shown that the luminosity of nuclear
star clusters in early- and intermediate type spirals correlates
with the total luminosity of the host galaxy. A similar result
has been established by \cite{lot01} for dE galaxies.
In Figure~\ref{fig:lumcorrs}a, we demonstrate that such a 
correlation is also found in spirals at the late end of the Hubble 
sequence. Here, we plot the apparent I-band magnitude of the
NCs, \mappic\ (from Paper~I), against the total apparent 
blue magnitude, \mappbg , of the host galaxy \citep*{pat94} as listed in the 
Lyon-Meudon Extragalactic Database\footnote{http://leda.univ-lyon1.fr/} (LEDA).
The relation still holds, albeit with a somewhat larger scatter, when
plotting the absolute magnitudes \mic\ and \mbg\ (Fig.~\ref{fig:lumcorrs}b). 
The conversion from apparent to absolute magnitudes and the
correction for Galactic reddening was performed using the foreground 
extinction and galaxy distances listed in Table~1 of Paper~I.
The solid lines in both panels indicate the best (least-squares) linear fit 
to the datapoints. The fit coefficients as well as their statistical 
significance (as measured by the Spearman rank-order coefficient) 
are listed in Table~\ref{tbl:coeff}. The fact that despite the large 
intrinsic variations of the cluster luminosities discussed in 
\S\ref{subsec:brightbig}, there still is a rather good correlation with 
galaxy luminosity suggests that the global properties of the host galaxy 
play a significant role for the formation processes of NCs.

\begin{figure}[htb]
\centerline{\epsfig{file=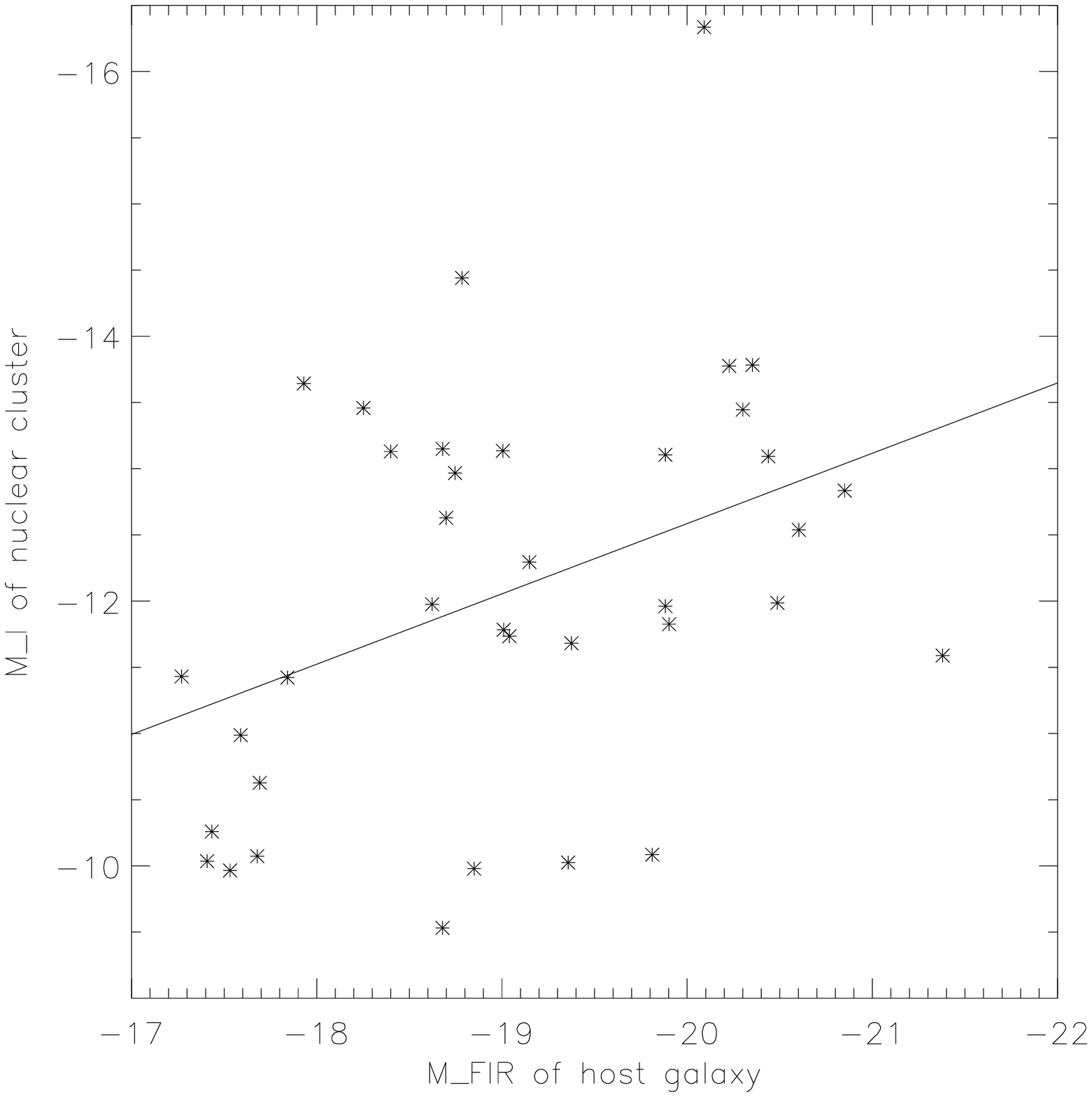, width=9cm} }
\caption[f8.eps]{\label{fig:fir}
	 d) I-band magnitude of NC vs. IRAS FIR
	 magnitude of the host galaxy (from the LEDA database).
	 Over the luminosity range of our galaxy sample, only
	 a weak correlation exists. The lines mark the best
	 least-squares linear fit to the data. The fit coefficients
	 and significance levels are given in Table~\ref{tbl:coeff}. }
\end{figure}

One possible mechanism that could link the occurrence of nuclear starbursts
to the total galaxy luminosity are tidal interactions 
or minor mergers, because these can both enhance the global
star formation rate (SFR) and cause significant infall of matter
towards the nucleus via re-distribution of angular momentum. 
However, as demonstrated in Fig.~\ref{fig:fir}, NC luminosity 
appears only weakly (if at all) correlated with the {\it IRAS} 
far-infrared (FIR) magnitude \mfir \footnote{The apparent far-infrared
magnitude, taken from LEDA, is calculated according to
$\rm m_{FIR} = -2.5 log(2.58\times f_{60} + f_{100}) + 14.75 $ where
$\rm f_{60}$ and $\rm f_{100}$ are the IRAS fluxes at $60\mum$ and $100\mum$
(in Jansky). This expression is identical to the one used in the RC3. The 
distances listed in Paper~I are then used to calculate \mfir .}
which is often used as a proxy for the SFR in galaxies 
\citep[see e.g.][for a review]{tel88}. Assuming that tidal interactions
contribute significantly to the SFR and hence the FIR luminosity in
late-type spirals, this lack of a strong correlation 
between SFR and NC luminosity suggests that tidal interactions are 
not the driving force for the evolution of the very nucleus. This is 
supported by the WFPC2 images of our galaxy sample (Paper~I) which 
show little evidence for recent star formation activity in their 
(central) disks. Instead, the smooth, shallow, and undisturbed 
stellar disks of many late-type spirals indicate a rather uneventful 
dynamical history, yet most of them have a luminous NC.  

An alternative explanation for the strong correlation between
\mic\ and \mbg\ could be that the properties of the NC
are more governed by galaxy mass than by the global star 
formation rate. In principle, HI linewidths could be used to address this
possibility: the inclination-corrected rotational component of the HI
linewidth
\beq 
	W_{\rm R}^i = (W_{20} - W_{\rm rand})/{\rm sin}\,i \ ,
\eeq
based on the correction for random motions of \cite{bot83},
can be used as a proxy for the dynamical mass
of the galaxy. Unfortunately, 
our galaxy sample was selected to have
inclinations close to face-on and consequently the $\sin\,i$ corrections are
highly uncertain. For the sake of completeness, we have netherless calculated
the galaxy masses with the above expression. Following \cite*{mat98}, 
we adopted $W_{\rm rand} = 20\kms$ for our galaxies, which corresponds to a 
line-of-sight velocity dispersion $\sigma_z = 5.5\kms$ \citep{rhe96}.
However, using the inclination $i$ as listed in LEDA, we found that
for our sample, the HI width does not even correlate with total galaxy
luminosity which is a contradiction of the Tully-Fisher relation.
We thus conclude that for our sample of face-on spirals, $W_{\rm R}^i$ is 
not a reliable indicator of galaxy mass, and that we therefore cannot 
draw any meaningful conclusions about possible correlations of NC
properties and host galaxy mass.

%%%%%%%%%%%%%%%%%%%%%%%%%%%%%%%%%%%%%%%%%%%%%%%%%%%%%%%%%%%%%%%%%%%%%%
\subsection{Central Surface Brightness of the Galaxy Disk}
\begin{figure}[htb]
\centerline{\epsfig{file=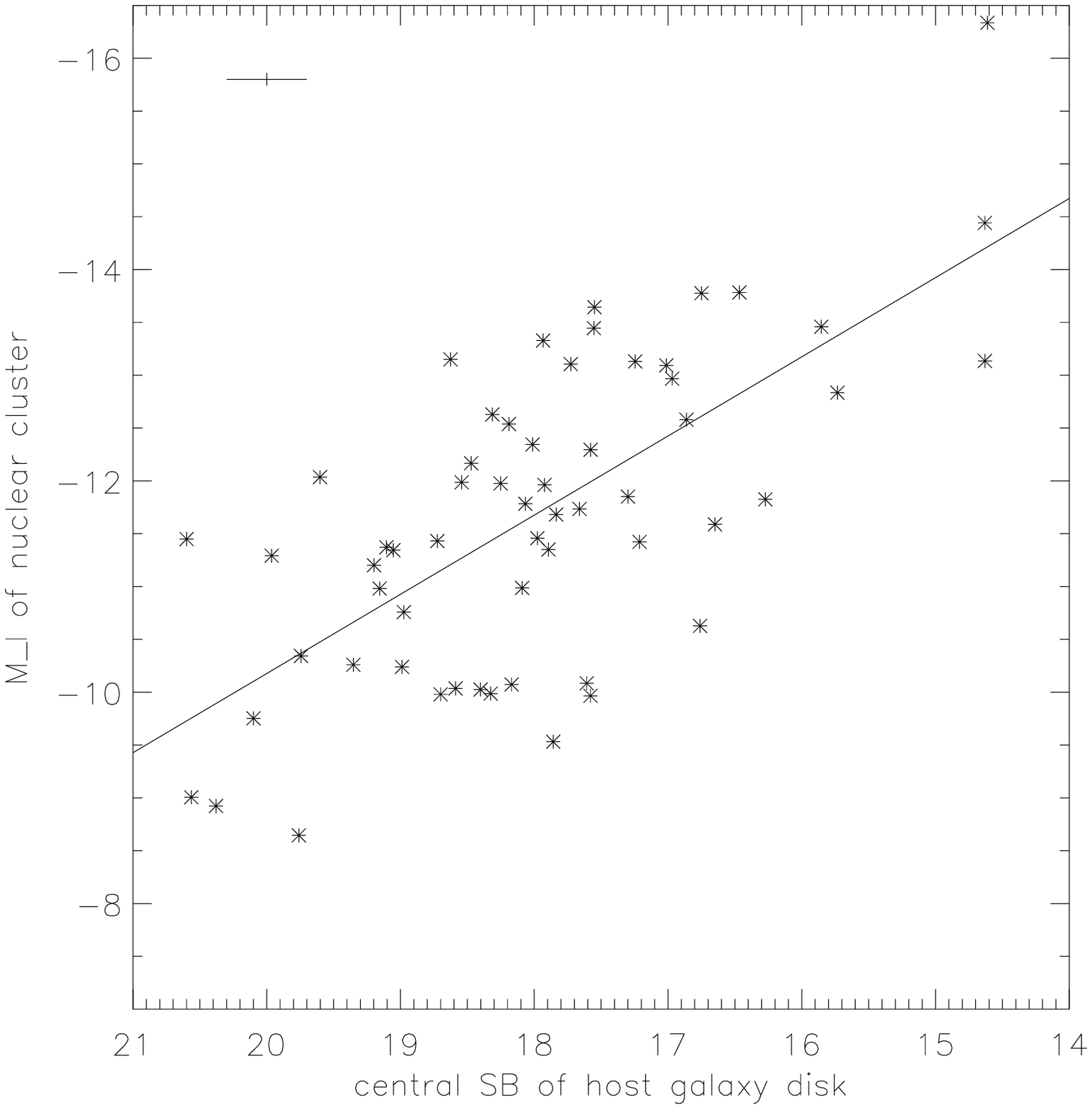, width=9cm}
	    \epsfig{file=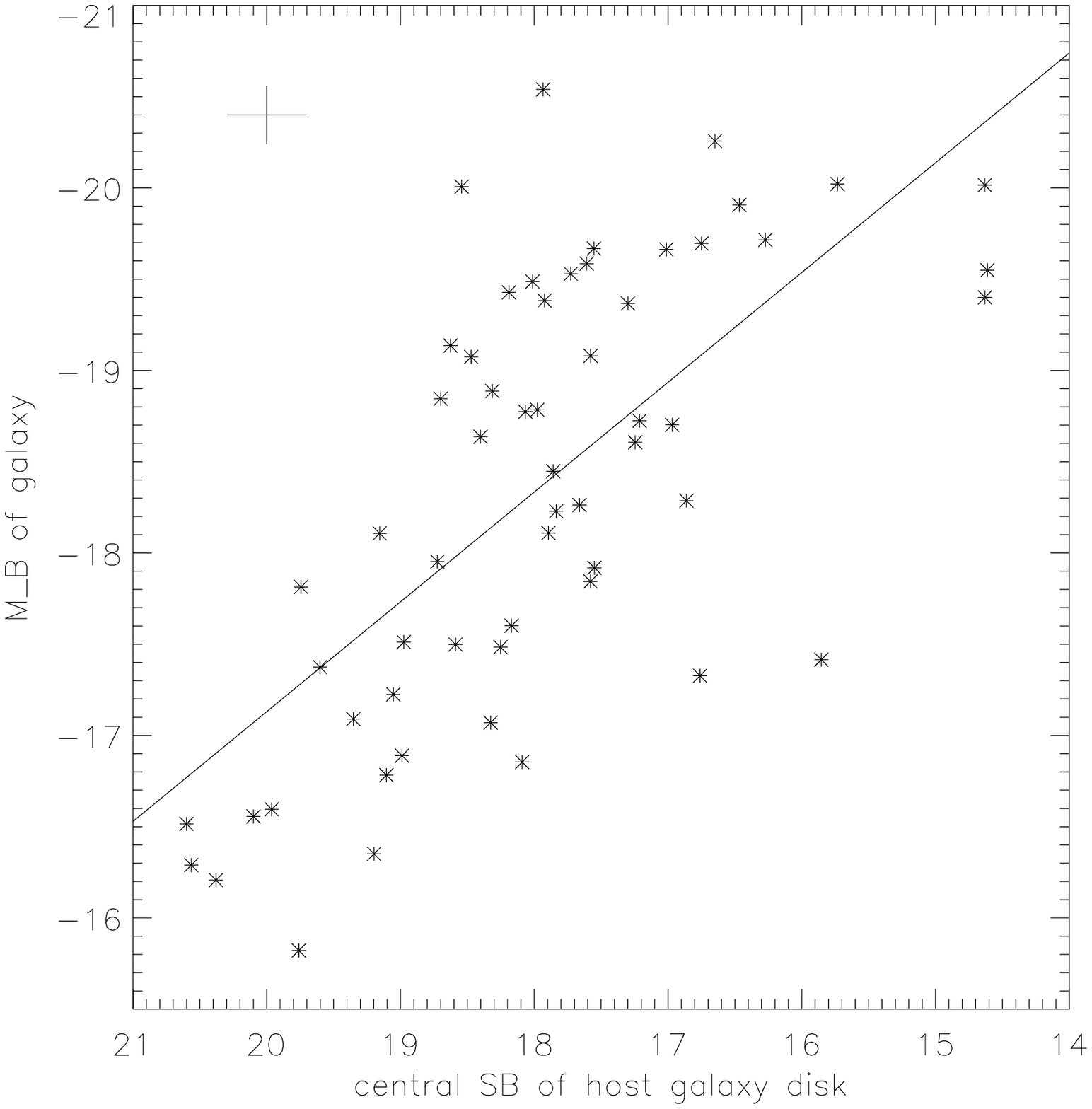, width=9cm} }
\caption[f9.eps]{\label{fig:sbcorr}
	 a) I-band magnitude of nuclear star cluster, \mic , and
	 b) total blue galaxy magnitude \mbg , as a function of
	 central disk surface brightness \mucd . All three quantities
	 are correlated with each other.}
\end{figure}

%%%%%%%%%%%%%%%%%%%%%%%%%%%%%%%%%%%%%%%%%%%%%%%%%%%%%%%%%%%%%%%%%%%%%%
Recent analysis of large galaxy samples \citep[e.g.][]{bla02}
has put on a solid statistical ground the notion that there are
strong correlations between a number of physical properties of
galaxies. In particular, \cite{bla02} have shown that for 
disk-dominated, ``exponential'' galaxies (i.e., with \cite{ser68} 
shape parameter $n < 1.5$), 
surface brightness correlates with total luminosity.
We confirm this finding for our sample, as demonstrated in 
Figure~\ref{fig:sbcorr}a. Here, we have plotted \mbg , the total
blue magnitude of the host galaxy, versus \mucd , the central 
I-band surface brightness of the stellar disk underlying the NC.
The values for \mucd\ are listed in Column~3 of Table~\ref{tbl:props};
they were measured from our WFPC2 images by averaging the two 
inward extrapolations of the surface brightness profile shown 
in Fig.~3 of Paper~I at a radius of $0.024\as$ (i.e. the border 
of the central WFPC2 pixel). 

Given that NC luminosity \mic\ correlates with host 
galaxy luminosity (Fig.~\ref{fig:lumcorrs}b), and that the latter correlates 
with \mucd , it is not surprising that \mic\ is also correlated with \mucd ,
as demonstrated in Figure~\ref{fig:sbcorr}b. Again, the fit coefficients and 
significance levels of both correlations are listed in Table~\ref{tbl:coeff}. 
While it is true in general that higher surface brightness disks make it more 
difficult to detect faint clusters, we emphasize that this effect does not 
result in a significant selection bias. In fact, most galaxies without clear
identification of a NC have \mucd\ values {\it below average}
(Figure~4 of Paper~I). Note that the values of \mucd\ are as observed,
i.e. they have not been corrected for inclination.

\begin{figure}[htb]
\centerline{\epsfig{file=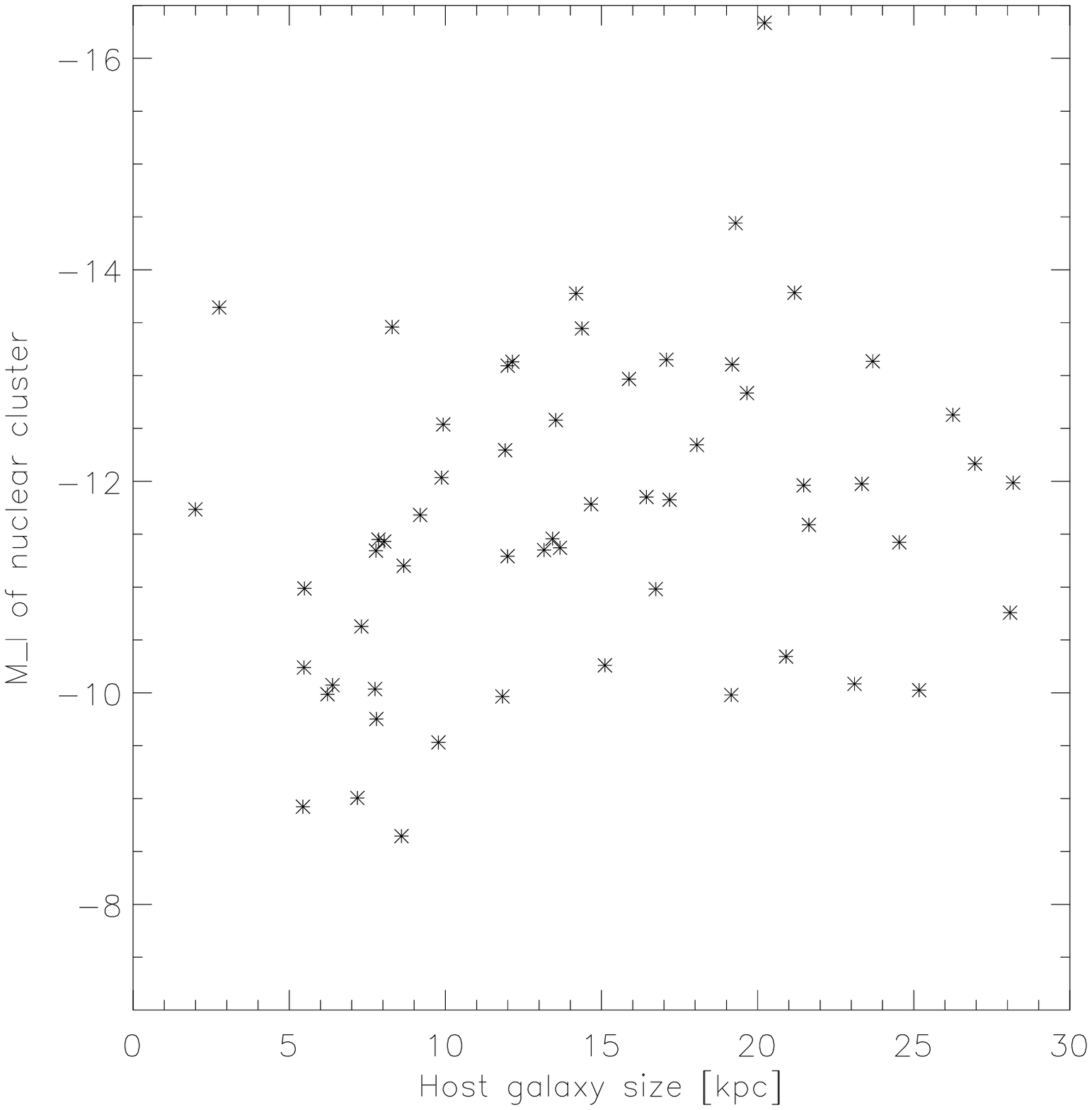, width=9cm}
	    \epsfig{file=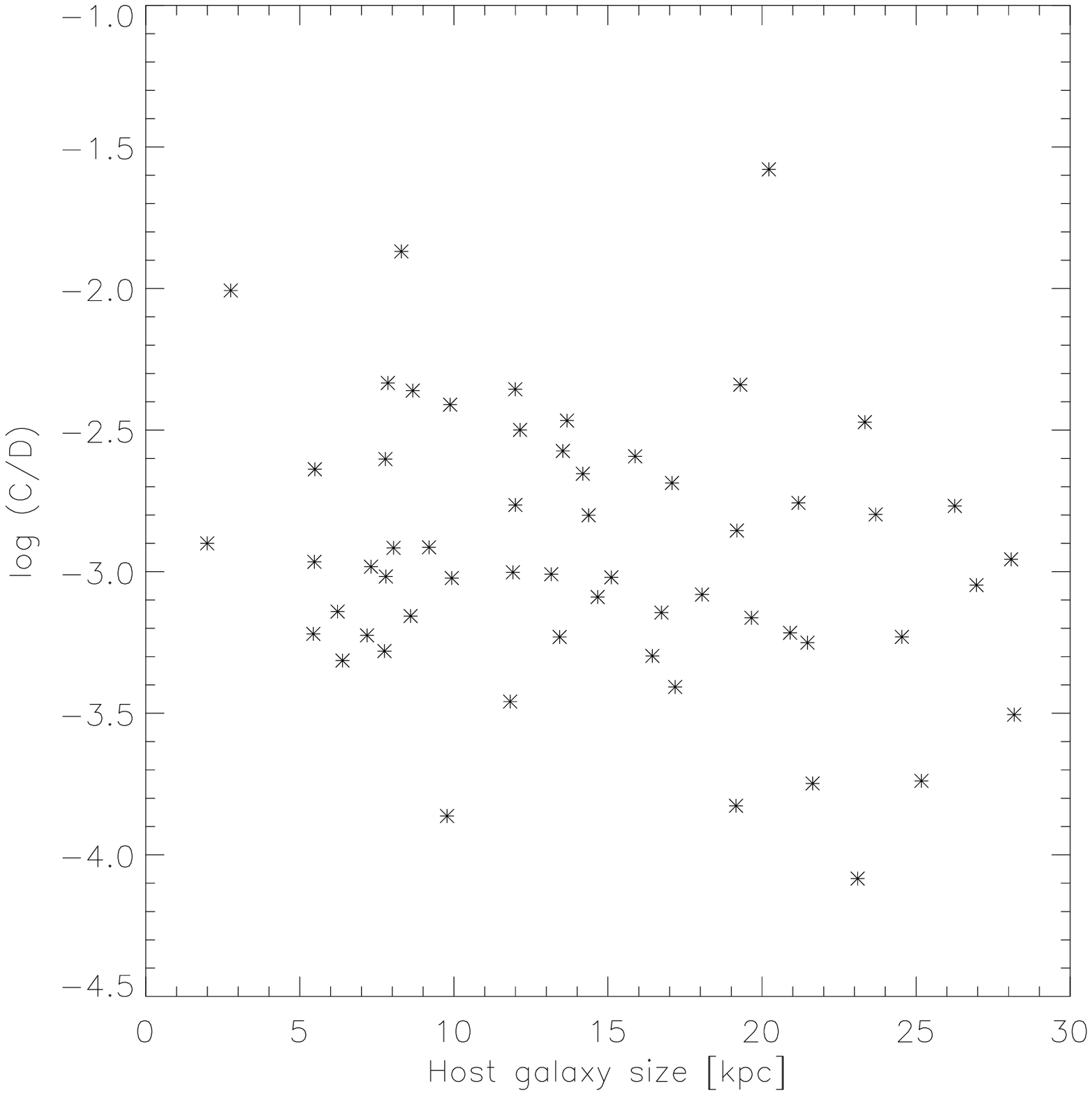, width=9cm} }
\caption[f10.eps]{\label{fig:size}
	 a) I-band magnitude of nuclear star cluster, \mic , and
	 b) cluster-to-disk luminosity ratio $C/D$ as a function
	 of physical isophotal diameter of the galaxy disk. Any 
	 apparent trend can be explained by the strong correlation
	 between cluster luminosity and galaxy luminosity (see text).}
\end{figure}

So far, we have shown that more luminous galaxies have higher surface 
brightness disks as well as more luminous NCs.
Is this simply a reflection of the popular notion that 
``bigger galaxies have more of everything''? While this is certainly
part of the answer, it is likely not the whole story. This is
demonstrated in Fig.~\ref{fig:size}, which plots \mic, as well as the
cluster-to-disk luminosity ratio $C/D$, as functions of the physical
size of the galaxy disk, i.e., the major axis diameter of the $\rm \mu_B=25$ 
isophote as listed in LEDA.
(The disk luminosity is calculated as total galaxy luminosity
minus NC luminosity, but with $C/D \la 0.01$ in general the disk light is not
significantly different from the total galaxy luminosity.)

Apparently, \mic\ depends weakly on galaxy size: the Spearman rank-order
correlation coefficient in Fig.~\ref{fig:size}a is $s=0.36$, with slightly
less than 3-$\sigma$ significance. However, this actually reflects the
stronger and more significant correlation between NC luminosity and galaxy
luminosity (\S4.1), coupled with a similarly significant dependence of disk
size on total disk luminosity: $R_{\rm disk}\sim L_{\rm disk}^{0.32}$ in this
sample. From Table \ref{tbl:coeff} we have that NC luminosity scales
with galaxy (disk) luminosity as $L_{\rm cl}\sim L_{\rm disk}^{0.78}$, and
thus we can expect $C/D\equiv (L_{\rm cl}/L_{\rm disk}) \sim
L_{\rm disk}^{-0.22} \sim R_{\rm disk}^{-0.69}$ even if there is no
{\it direct} connection between NC luminosity and global disk size. This
scaling is indeed consistent with the weak trend in Fig.~\ref{fig:size}b.

To first order, the NC luminosity is a measure of the efficiency with
which gas is funneled towards the nucleus and converted into
stars, regardless of the exact star formation history of the NC.
This is true even if the cluster luminosity is dominated by a minor
(in mass) but recent burst, because on average, younger populations imply 
a higher duty cycle of nuclear ``delta''-bursts. It thus appears that, 
at least to some degree, the mechanisms that regulate the (past) star
formation  efficiency in the vicinity of the galaxy nucleus -- as traced by
the central surface brightness of its stellar disk -- also play a significant
role for the amount of gas that is funneled into the very nucleus and
converted into stars. Further support for this notion comes from the fact
that the molecular gas content within the central kpc of late-type spirals
also appears to be connected to the disk surface brightness in the sense that
low surface brightness disks in general have little or no central molecular
gas \citep*[Fig.~7 of][]{boe03b}.

%
%%%%%%%%%%%%%%%%%%%%%%%%%%%%%%%%%%%%%%%%%%%%%%%%%%%%%%%%%%%%%%%%%%%%%%
\subsection{Hubble-type}
\begin{figure}[htb]
\centerline{\epsfig{file=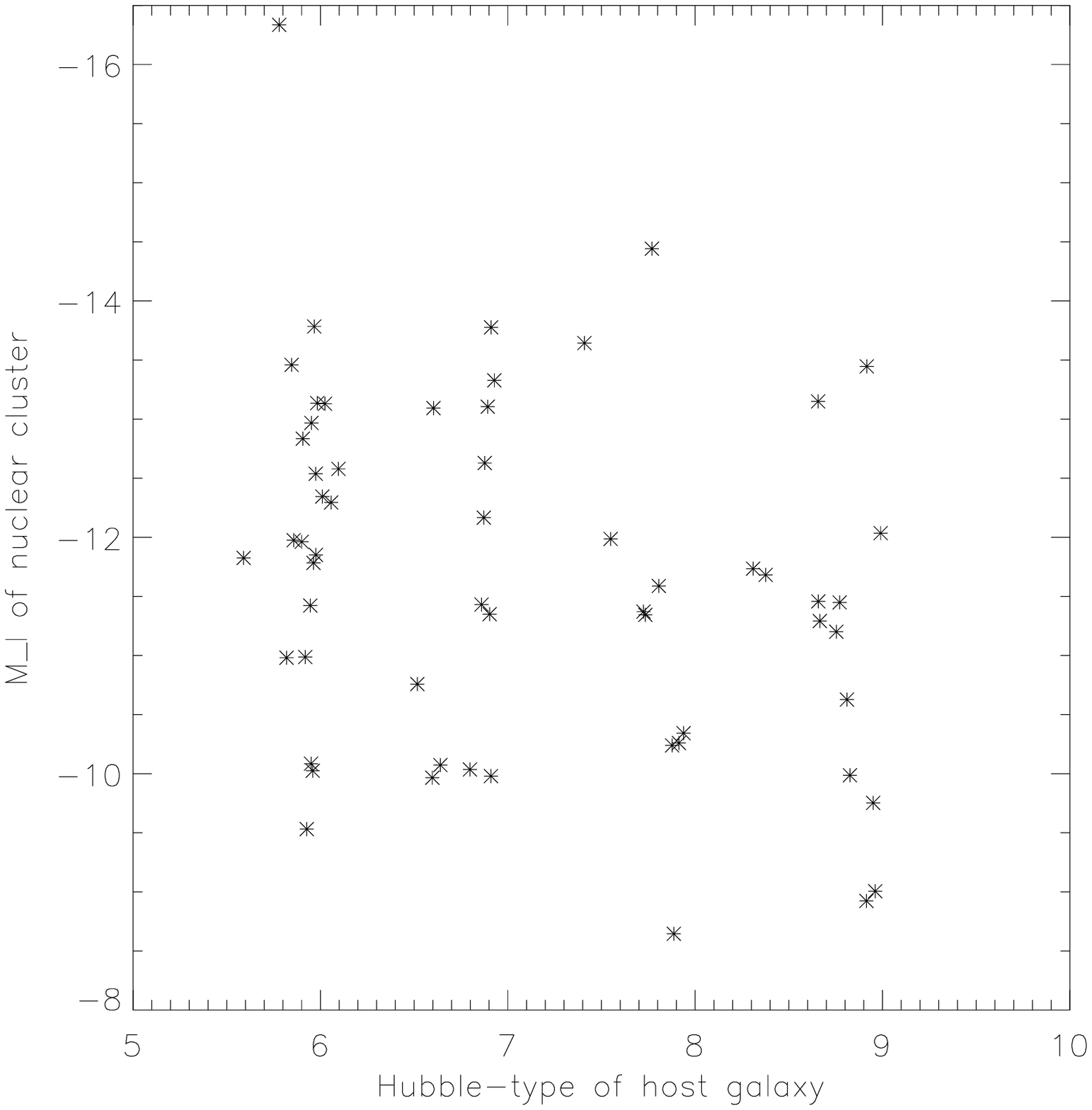, width=9cm}
	    \epsfig{file=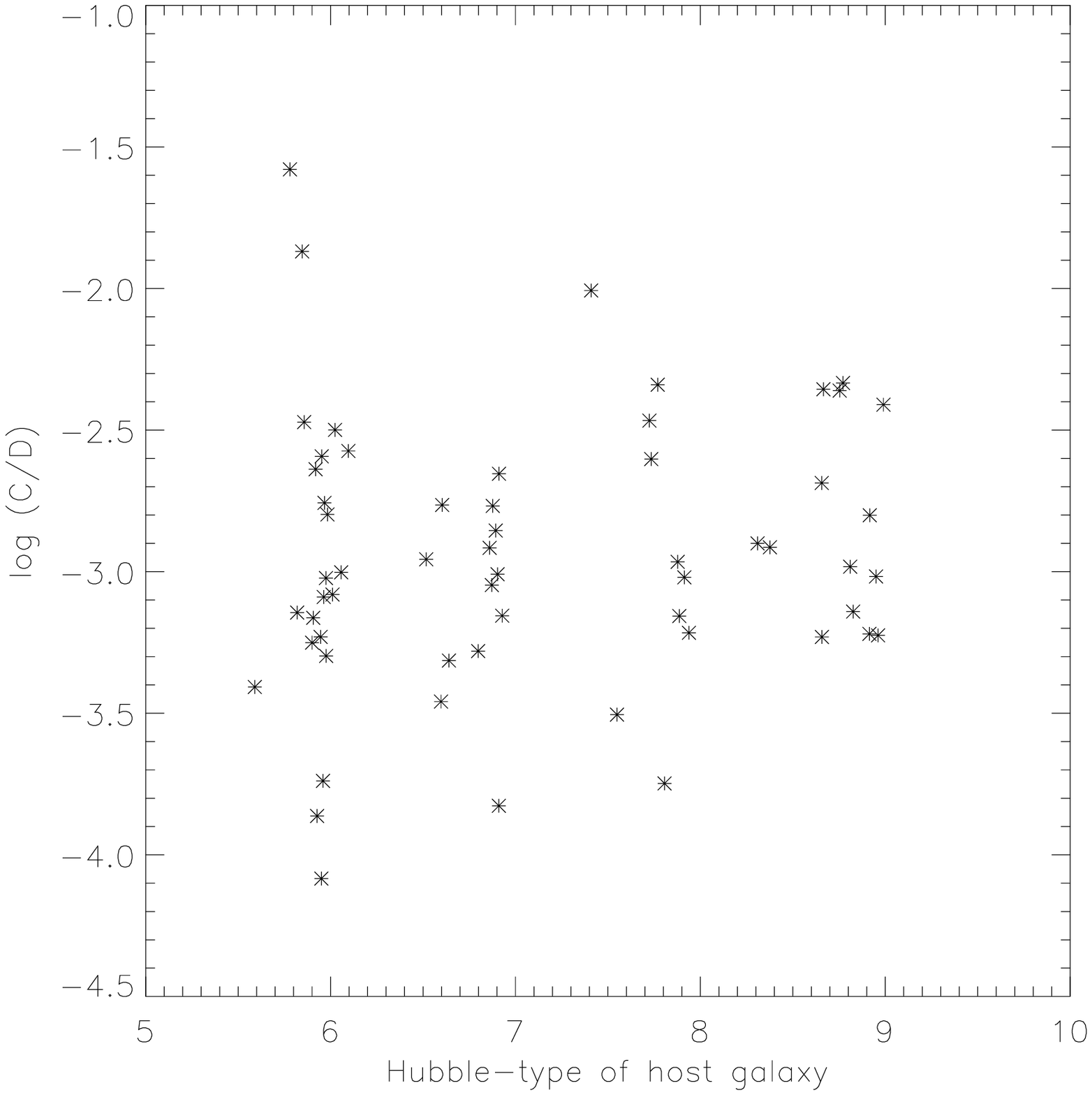, width=9cm}
	    }
\caption[f11.eps]{\label{fig:htype}
	 a) I-band magnitude of nuclear star cluster, \mic , and
	 b) cluster-to-disk luminosity ratio $C/D$ as a function of
	galaxy Hubble type. No trends are apparent.}
\end{figure}

%%%%%%%%%%%%%%%%%%%%%%%%%%%%%%%%%%%%%%%%%%%%%%%%%%%%%%%%%%%%%%%%%%%%%%
Figure~\ref{fig:htype}a compares nuclear cluster absolute magnitudes
against the Hubble type of their host galaxies. The very weak correlation
apparent here, with a Spearman coefficient
$s=-0.26$, is only marginally significant (at the 2-$\sigma$ level, and less
if the cluster in NGC 7418, with $M_I^C=-16.33$, is excluded). It seems in
any case to be the result of the strong link between NC and total galaxy
luminosities, together with a trend towards slightly fainter galaxy
magnitudes, on average, for the later Hubble types. Figure~\ref{fig:htype}b
therefore plots the ratio of cluster-to-disk luminosity
vs.~Hubble type. The correlation coefficient in this case is only $s=0.13$,
which is not significant at even the 1-$\sigma$ level.

Thus, nuclear cluster luminosity is essentially independent of Hubble type.
This should not come as a surprise given that the Hubble type
is not a strong discriminant for the latest-type spirals, as discussed 
in more detail by \cite*{boe03}. The main reason is that the NC 
is quite easily mistaken for a compact stellar bulge in
seeing-limited images, and hence a ground-based morphological classification 
based at least in part on the bulge-to-disk ratio is inaccurate at best.

%
%%%%%%%%%%%%%%%%%%%%%%%%%%%%%%%%%%%%%%%%%%%%%%%%%%%%%%%%%%%%%%%%%%%%%%
\subsection{Nuclear Clusters and Stellar Bars}
%%%%%%%%%%%%%%%%%%%%%%%%%%%%%%%%%%%%%%%%%%%%%%%%%%%%%%%%%%%%%%%%%%%%%%
Using the morphological classifications of our galaxy
sample, both from the RC3 (listed in Table~1 of Paper~I)
and the LEDA database, we have investigated for our complete Paper I
sample whether nuclear star clusters are more likely to be found in 
(or whether they preferentially avoid) galaxies of a certain 
bar class. Table~\ref{tbl:bars} summarizes the statistics for 
both subsamples with (59 galaxies) and without (18 galaxies)
a NC. The fraction of galaxies with a NC is
76\%, 88\%, and 67\% for the RC3 classifications A, AB, and B, 
respectively. The LEDA database only lists the types ``unbarred'' and
``barred''; for these, the corresponding fractions are
88\% and 73\%. The apparently smaller frequency of NCs in
barred galaxies is probably not significant, given the small
number statistics in some categories, and the rather high number 
of uncertain classifications (17 galaxies in our sample are classified 
either as uncertain or peculiar in the RC3). More reliable
classifications of the morphology of late-type spirals will require 
a more accurate isophotal analysis or even kinematic information on their
central disks.

This rough analysis suggests that the galaxy bar class is not
a significant factor for the likelihood of hosting a NC. In particular,
the late-type galaxies in our sample seem to have no especial difficulty
in supporting bars and nuclear clusters simultaneously. This might have
seemed somewhat paradoxical in the light of some early numerical simulations
of barred galaxies, which found that fairly modest central mass
concentrations could drive dynamical instabilities that destroyed the bar
on rather short timescales \citep*[e.g.,][]{frie94,nor96}. However,
more recent work \citep{shen03} suggests that even an NC as massive as a few
percent of the total disk mass of a galaxy (i.e., with $C/D\ga 0.02$, already
much higher than the typical values in Fig.~\ref{fig:size}b) may not be
able to destroy a bar completely within a Hubble time. It is then somewhat
unclear what role bars and NCs together might play in the formation of the
exponential ``pseudo''-bulges found in many late-type spirals
\citep*[cf.][]{car01}.

%%%%%%%%%%%%%%%%%%%%%%%%%%%%%%%%%%%%%%%%%%%%%%%%%%%%%%%%%%%%%%%%%%%%%%
\section{Summary} 
\label{sec:summary}
%%%%%%%%%%%%%%%%%%%%%%%%%%%%%%%%%%%%%%%%%%%%%%%%%%%%%%%%%%%%%%%%%%%%%%
We have presented a structural analysis of 39 nuclear
star clusters in late-type spiral galaxies. After correction
for the instrumental point spread function, we measure
a median effective radius for the sample of 
$\overline{r_{\rm e}}=3.5$ pc, with 50\% 
of the sample falling between $2.4\ {\rm pc}\le r_{\rm e}\le 5.0\ {\rm pc}$. 
This narrow size distribution is statistically indistinguishable from 
that of Galactic globular clusters, even though the nuclear clusters 
are on average 4 magnitudes brighter than the old globulars.
The compactness of NCs clearly separates them from larger
ellipsoidal stellar systems such as spiral bulges or elliptical
galaxies.

In addition, we have compared the luminosity of NCs with 
various properties of their host galaxies. We find that
NC luminosity is correlated with galaxy luminosity which extends
a previous result obtained for earlier-type spirals. 
Because there is a strong correlation between total
luminosity and central disk surface brightness for the galaxies in
our sample, NC luminosity also correlates with local disk brightness. 
It is not clear which of these trends is physically more fundamental. 
Either way, other apparent correlations such as between cluster
luminosity and galaxy size can ultimately be explained by the underlying
link to galaxy luminosity and/or central disk surface brightness.

We find no systematic
trends with the Hubble type of the galaxy, confirming that the latter is
not a strong discriminant for the latest-type spirals. We also find
no evidence for a connection between the presence/absence of NCs
and the presence/absence of large-scale stellar bars in the host
galaxy.

Without additional spectral information, it is not 
possible to put strong constraints on the formation history
of NCs. We suggest that the most likely scenario for
NC formation consists of multiple nuclear starbursts, possibly
caused by gas infall onto a ``seed'' cluster. However, a more
detailed discussion of this hypothesis has to await the analysis 
of our spectroscopic data which we will present in a future paper.

\acknowledgements
We are grateful to S. Larsen for help with the implementation
and use of his \ish\ software package.
Support for this work was provided by the National Aeronautics and
Space Administration (NASA) through grants for project no. GO-8599
awarded by the Space Telescope Science Institute, which is operated by
AURA, Inc., under NASA contract no. NAS5-26555.
This research has made use of the NASA/IPAC Extragalactic Database (NED) 
which is operated by the Jet Propulsion Laboratory, California Institute of 
Technology, under contract with NASA. 
It has also benefited greatly from use of the Lyon-Meudon
Extragalactic Database (LEDA, http://leda.univ-lyon1.fr).
\newpage
\begin{deluxetable}{lcccccccc}
\tabletypesize{\scriptsize}
\tablecaption{Summary of Cluster Properties \label{tbl:props}}
\tablewidth{0pt}
\tablehead{
\colhead{(1)} & \colhead{(2)} & \colhead{(3)} & \colhead{(4)} &
\colhead{(5)} & \colhead{(6)} & \colhead{(7)} & \colhead{(8)} &
\colhead{(9)} \\
\colhead{Galaxy} & \colhead{Distance} & \colhead{Disk $\mu_I^0$} &
\colhead{$m_I^C$}  & \colhead{$M_I^C$} & \colhead{Best Model} &
\colhead{$r_{\rm e}$} & \colhead{$r_{\rm e}$} &
\colhead{${\rm log}(I_{\rm e})$} \\
 & \colhead{[Mpc]} & \colhead{[mag/arcsec$^2$]} &  \colhead{[mag]} &
\colhead{[mag]} &  & \colhead{[arcsec]} & \colhead{[pc]} &
\colhead{[$\lsun$/pc$^2$]} 
}
\startdata
 A1156+52  & 18.7 & 19.21 & 20.43 & $-10.98$ &  King100 & 0.025$\pm$0.006 &  2.27 & 4.52 \\
 ESO138-10 & 13.5 & 15.06 & 16.64 & $-14.44$ &   King15 & 0.104$\pm$0.041 &  6.81 & 4.94 \\
 ESO202-41 & 19.9 & 20.58 & 22.51 &  $-9.01$ &  ngf$^1$ &      $-$        &   $-$ &  $-$ \\
 ESO241-6  & 17.4 & 18.35 & 21.24 &  $-9.99$ &  ngf$^1$ &      $-$        &   $-$ &  $-$ \\
 ESO290-39 & 19.1 & 20.41 & 22.52 &  $-8.92$ &  ngf$^1$ &      $-$        &   $-$ &  $-$ \\
 ESO358-5  & 20.1 & 20.62 & 20.09 & $-11.45$ &  King100 & 0.046$\pm$0.004 &  4.48 & 4.11 \\
 ESO418-8  & 14.1 & 19.02 & 20.54 & $-10.24$ & Moffat15 & 0.019$\pm$0.005 &  1.30 & 4.70 \\
 ESO504-30 & 23.9 & 19.20 & 20.70 & $-11.34$ &  King100 & 0.047$\pm$0.012 &  5.45 & 3.90 \\
 NGC0275   & 24.0 & 18.30 & 19.47 & $-12.54$ &  King100 & 0.030$\pm$0.004 &  3.49 & 4.76 \\
 NGC0300   &  2.2 & 18.75 & 15.29 & $-11.43$ &   King30 & 0.272$\pm$0.066 &  2.90 & 4.48 \\
 NGC0337a  & 14.3 & 19.93 & 20.62 & $-10.34$ &  ngf$^1$ &      $-$        &   $-$ &  $-$ \\
 NGC0428   & 16.1 & 18.68 & 17.94 & $-13.15$ &  King100 & 0.043$\pm$0.002 &  3.36 & 5.04 \\
 NGC0450   & 25.6 & 18.00 & 20.07 & $-11.96$ &  King100 & 0.196$\pm$0.050 & 24.33 & 2.85 \\
 NGC0600   & 25.2 & 18.54 & 19.91 & $-12.17$ &  King100 & 0.024$\pm$0.003 &  2.93 & 4.77 \\
 NGC0853   & 20.2 & 17.89 & 19.89 & $-11.68$ &   King30 & 0.035$\pm$0.004 &  3.43 & 4.44 \\
 NGC1042   & 18.2 & 14.69 & 18.22 & $-13.14$ &  King100 & 0.022$\pm$0.005 &  1.94 & 5.51 \\
 NGC1493   & 11.4 & 17.27 & 17.17 & $-13.13$ & Moffat15 & 0.047$\pm$0.005 &  2.60 & 5.26 \\
 NGC2139   & 23.6 & 15.80 & 19.09 & $-12.83$ &  ngf$^2$ &      $-$        &   $-$ & $-$ \\
 NGC2552   &  9.9 & 19.69 & 18.04 & $-12.04$ &  King100 & 0.046$\pm$0.004 &  2.21 & 4.96 \\
 NGC2763   & 25.3 & 16.41 & 20.33 & $-11.83$ & Moffat25 & 0.019$\pm$0.004 &  2.33 & 4.83 \\
 NGC2805   & 28.1 & 18.03 & 19.02 & $-13.33$ & Moffat15 & 0.056$\pm$0.003 &  7.63 & 4.40 \\
 NGC3346   & 18.8 & 18.12 & 19.64 & $-11.78$ &  King100 & 0.020$\pm$0.003 &  1.82 & 5.03 \\
 NGC3423   & 14.6 & 17.36 & 19.03 & $-11.85$ &  King100 & 0.059$\pm$0.008 &  4.18 & 4.33 \\
 NGC3445   & 32.1 & 17.57 & 19.10 & $-13.45$ &  King100 & 0.024$\pm$0.005 &  3.73 & 5.07 \\
 NGC3782   & 13.5 & 18.20 & 20.61 & $-10.07$ & Moffat15 & 0.029$\pm$0.007 &  1.90 & 4.31 \\
 NGC3906   & 16.7 & 18.64 & 21.12 & $-10.04$ &  King100 & 0.047$\pm$0.013 &  3.81 & 3.69 \\
 NGC3913   & 17.0 & 17.60 & 21.21 &  $-9.97$ &  ngf$^2$ &      $-$        &   $-$ &  $-$ \\
 NGC4027   & 22.7 & 16.73 & 20.27 & $-11.59$ &  ngf$^2$ &      $-$        &   $-$ &  $-$ \\
 NGC4204   & 13.8 & 19.42 & 20.51 & $-10.26$ &  King100 & 0.052$\pm$0.011 &  3.48 & 3.85 \\
 NGC4299   & 16.8 & 17.73 & 19.46 & $-11.73$ &  King100 & 0.016$\pm$0.004 &  1.30 & 5.30 \\
 NGC4411b  & 19.1 & 16.92 & 18.88 & $-12.58$ &  King100 & 0.097$\pm$0.007 &  8.98 & 3.96 \\
 NGC4416   & 20.7 & 17.91 & 22.10 &  $-9.53$ &  ngf$^1$ &      $-$        &   $-$ &  $-$ \\
 NGC4487   & 14.6 & 17.01 & 17.89 & $-12.97$ & Moffat15 & 0.010$\pm$0.003 &  0.71 & 6.32 \\
 NGC4496a  & 25.3 & 18.59 & 20.08 & $-11.99$ & Moffat15 & 0.020$\pm$0.003 &  2.45 & 4.85 \\
 NGC4540   & 19.8 & 17.64 & 19.25 & $-12.29$ &  ngf$^2$ &      $-$        &   $-$ &  $-$ \\
 NGC4618   & 10.7 & 18.02 & 18.73 & $-11.46$ &  ngf$^2$ &      $-$        &   $-$ &  $-$ \\
 NGC4625   & 11.7 & 16.80 & 19.74 & $-10.63$ &  King100 & 0.508$\pm$0.040 & 28.81 & 2.17 \\
 NGC4701   & 11.0 & 15.91 & 16.80 & $-13.46$ &  King100 & 0.046$\pm$0.001 &  2.45 & 5.44 \\
 NGC4775   & 22.4 & 16.82 & 18.04 & $-13.78$ &  King100 & 0.039$\pm$0.004 &  4.24 & 5.09 \\
 NGC5068   &  8.7 & 18.21 & 17.54 & $-12.34$ &   King30 & 0.177$\pm$0.064 &  7.47 & 4.03 \\
 NGC5584   & 24.2 & 17.68 & 21.91 & $-10.08$ &  ngf$^1$ &      $-$        &   $-$ &  $-$ \\
 NGC5585   &  8.2 & 17.92 & 18.24 & $-11.35$ &  King100 & 0.082$\pm$0.006 &  3.26 & 4.35 \\
 NGC5668   & 23.8 & 17.80 & 18.85 & $-13.10$ &  King100 & 0.038$\pm$0.004 &  4.38 & 4.79 \\
 NGC5669   & 21.2 & 18.45 & 21.65 & $-10.03$ &  ngf$^1$ &      $-$        &   $-$ &  $-$ \\
 NGC5774   & 23.5 & 18.78 & 21.96 &  $-9.98$ &  ngf$^1$ &      $-$        &   $-$ &  $-$ \\
 NGC5964   & 22.2 & 18.43 & 19.21 & $-12.63$ &  King100 & 0.046$\pm$0.003 &  4.95 & 4.50 \\
 NGC6509   & 27.5 & 17.39 & 19.48 & $-13.09$ &  King100 & 0.021$\pm$0.005 &  2.80 & 5.18 \\
 NGC7418   & 18.4 & 14.64 & 15.02 & $-16.34$ &  ngf$^2$ &      $-$        &   $-$ &  $-$ \\
 NGC7424   & 10.9 & 17.23 & 18.79 & $-11.42$ &  ngf$^2$ &      $-$        &   $-$ &  $-$ \\
 NGC7689   & 24.9 & 16.49 & 18.22 & $-13.78$ &  ngf$^2$ &      $-$        &   $-$ &  $-$ \\
 NGC7793   &  3.3 & 17.59 & 13.99 & $-13.64$ &  King100 & 0.484$\pm$0.006 &  7.74 & 4.51 \\
 UGC3574   & 23.4 & 18.35 & 19.97 & $-11.98$ &  King100 & 0.068$\pm$0.008 &  7.71 & 3.85 \\
 UGC3826   & 27.8 & 19.11 & 21.59 & $-10.76$ &  ngf$^1$ &      $-$        &   $-$ &  $-$ \\
 UGC4499   & 12.5 & 19.83 & 21.91 &  $-8.65$ &  ngf$^1$ &      $-$        &   $-$ &  $-$ \\
 UGC4988   & 24.2 & 19.23 & 20.76 & $-11.20$ &  King100 & 0.030$\pm$0.006 &  3.52 & 4.22 \\
 UGC5015   & 25.7 & 19.14 & 20.71 & $-11.37$ &   King30 & 0.060$\pm$0.017 &  7.48 & 3.64 \\
 UGC6931   & 20.7 & 20.15 & 21.87 &  $-9.75$ &  ngf$^1$ &      $-$        &   $-$ &  $-$ \\
 UGC8516   & 16.5 & 18.15 & 20.16 & $-10.99$ &  ngf$^2$ &      $-$        &   $-$ &  $-$ \\
 UGC12732  & 12.4 & 20.14 & 19.35 & $-11.29$ &  King100 & 0.065$\pm$0.003 &  3.91 & 4.17 \\
\enddata
\tablecomments{(2): galaxy distance from Paper~I.
(3): observed (i.e. not inclination-corrected) peak surface brightness 
of the galaxy disk (underlying the NC).
(4): apparent $I$-band magnitude of NC.
(5): absolute $I$-band magnitude of NC, corrected for
foreground reddening. 
(6):  name of best-fitting \ish\ model, ``ngf'' means that no good fit was
possible, either because of low signal-to-noise data (denoted with
superscript 1), or a complex morphology (superscript 2).
(7) angular effective radius and uncertainty as discussed in
\S~\ref{sec:sizes}.
(8) effective radius in parsec.
(9) logarithm of effective I-band intensity of NC, defined
as  $\log(I_{\rm e}) = 0.4(4.08-M_I^C)-\log(2\pi r_e^2)$ for $r_e$ in pc.
}
%\tablerefs{
%(a) \cite{ber92}
%}
\end{deluxetable}
%%%%%%%%%%%%%%%%%%%%%%%%%%%%%%%%%%%%%%%%%%%%%%%%%%%%%%%%%%%%%%%%%%%%%%%%

\clearpage

\begin{deluxetable}{lcccccc}
\tabletypesize{\scriptsize}
\tablecaption{Fit coefficients and statistical significance\label{tbl:coeff}}
\tablewidth{0pt}
\tablehead{
\colhead{(1)} & \colhead{(2)} & \colhead{(3)} & \colhead{(4)} &  
\colhead{(5)} & \colhead{(6)} & \colhead{(7)}\\
\colhead{x} & \colhead{y} & \colhead{Figure} & \colhead{a} & 
\colhead{b} & \colhead{SROCC} & \colhead{Sign. Level}
}
\startdata
\mappbg & \mappic & 6a &   6.8$\pm$1.5 & 0.99$\pm$0.11 & 0.757 & 5.762  \\
		  &    &     				  \\
\mbg    & \mic    & 6b &   2.6$\pm$2.3 & 0.78$\pm$0.13 & 0.654 & 4.978  \\
		  &    &     				  \\
\mfirg  & \mic    & 7  &  -2.0$\pm$3.7 & 0.53$\pm$0.20 & 0.420 & 2.554  \\
		  &    &     				  \\
\mucd   & \mbg    & 9a & -29.2$\pm$1.6 & 0.60$\pm$0.09 & 0.679 & 5.173  \\
		  &    &    				  \\
\mucd   & \mic    & 9b & -25.2$\pm$0.6 & 0.75$\pm$0.11 & 0.627 & 4.778  \\
		  &    &     				  \\
\enddata
\tablecomments{Cols. (1-3): quantities to be correlated using
the expression $y = a + b\times x$, and Figure in which they are 
plotted against each other. Cols. (4) \& (5): coefficients a and b of 
best linear least-squares fit, together with their statistical 
uncertainties (1$\sigma$). Cols. (6) \& (7): Spearman rank-order correlation
coefficient (SROCC) and significance (in $\sigma$) of correlation. }
\end{deluxetable}
%%%%%%%%%%%%%%%%%%%%%%%%%%%%%%%%%%%%%%%%%%%%%%%%%%%%%%%%%%%%%%%%%%%%%%%%
%%%%%%%%%%%%%%%%%%%%%%%%%%%%%%%%%%%%%%%%%%%%%%%%%%%%%%%%%%%%%%%%%%%%%%%%
%\newpage
%
\begin{deluxetable}{clcc}
\tabletypesize{\scriptsize}
\tablecaption{Impact of bar class on nuclear cluster frequency\label{tbl:bars}}
\tablewidth{0pt}
\tablehead{
\colhead{(1)} & \colhead{(2)} & \colhead{(3)} & \colhead{(4)}\\
\colhead{Database} & \colhead{Class} & \colhead{with NC} & \colhead{without NC} 
}
\startdata
	& unbarred (A)	& 76\% (16)  & 24\% (5)  \\
RC3	& mixed (AB)	& 88\% (23)  & 12\% (3)  \\
	& barred (B)	& 67\% (20)  & 33\% (10)  \\
	& total		& 77\% (59)  & 23\% (18)  \\
\hline
	& unbarred 	& 88\% (15)  & 12\% (2)  \\
LEDA	& barred 	& 73\% (44)  & 27\% (16)  \\
	& total		& 77\% (59)  & 23\% (18)  \\
\enddata
\tablecomments{Sample fraction and total number of galaxies with and without
a clearly identified nuclear cluster, sorted by bar classification. For
comparison, we have listed the statistics for both the RC3 and LEDA databases.}
\end{deluxetable}
%%%%%%%%%%%%%%%%%%%%%%%%%%%%%%%%%%%%%%%%%%%%%%%%%%%%%%%%%%%%%%%%%%%%%%%%
%%%%%%%%%%%%%%%%%%%%%%%%%%%%%%%%%%%%%%%%%%%%%%%%%%%%%%%%%%%%%%%%%%%%%%%%
\clearpage

\newpage

%\newpage

\newpage

%\newpage

%\newpage

%\newpage

%\newpage

%\newpage

%\newpage


\begin{thebibliography}{}
%\bibitem[Athanassoula (1992)]{ath92a}
%	Athanassoula, E. 1992, MNRAS, 259, 328
%\bibitem[Athanassoula(1992b)]{ath92b}
%        Athanassoula, E. 1992b, MNRAS, 259, 345
\bibitem[Barmby \ea (2002)Barmby, Holland, \& Huchra]{bar02}
	Barmby, P., Holland, S., \& Huchra, J. P. 2002, AJ, 123, 1937
\bibitem[Barth \ea (1995)]{barth95}
        Barth, A. J., Ho, L. C., Filippenko, A. V., \& Sargent, W. L.
        1995, AJ, 110, 1009
\bibitem[Bender \ea (1992)Bender, Burstein, \& Faber]{ben92}
	Bender, R., Burstein, D., \& Faber, S.M. 1992, ApJ, 399, 462
\bibitem[Blanton \ea (2002)]{bla02}
	Blanton, M. R. \ea\ 2002, ApJ, submitted (astro-ph/0209479)
\bibitem[B\"oker \ea (1999)B\"oker, van der Marel, \& Vacca]{boe99}
	B\"oker, T., van der Marel, R. P., \& Vacca, W. D. 1999, AJ, 118, 831
\bibitem[B\"oker \ea (2001)]{boe01}
	B\"oker, T., van der Marel, R. P., Mazzuca, L., Rix, H.-W.,
	Rudnick, G., Ho, L., \& Shields, J. C. 2001, AJ, 121, 1473
\bibitem[B\"oker \ea (2002)]{boe02}
	B\"oker, T., Laine, S., van der Marel, R. P., Sarzi, M.,
	Rix, H.-W., Ho, L., \& Shields, J. C. 2002, AJ, 123, 1389 (Paper I)
\bibitem[B\"oker \ea (2003)B\"oker, Stanek, \& van der Marel]{boe03}
	B\"oker, T., Stanek, R., \& van der Marel, R. P. 2003, AJ, 125, 1389
\bibitem[B\"oker \ea (2003b)B\"oker, Lisenfeld \& Schinnerer]{boe03b}
	B\"oker, T., Lisenfeld, \& U. Schinnerer, E. 2003, A\&A, 406, 87
\bibitem[Bottinelli \ea (1983)]{bot83}
	Bottinelli, L., Goughenheim, L., Paturel, G., \& de Vaucouleurs, G. 
	1983, A\&A, 118, 4
\bibitem[Burch \ea (1983)Burch, Gull, \& Skilling]{bur83}
	Burch, S.~F., Gull, S.~F., \& Skilling, J.\ 1983, Computer Vision 
	Graphics and Image Processing, 23, 113
\bibitem[Burstein \ea (1997)]{bur97}
	Burstein, D., Bender, R., Faber, S., \& Nolthenius, R. 1997, AJ, 114, 1365
%\bibitem[Byun \ea (1996)]{byu96}
%	Byun, Y.-I., et al. 1996, AJ, 111, 1889
\bibitem[Cardelli \ea (1989)Cardelli, Clayton \& Mathis]{car89}
        Cardelli, J. A., Clayton, G. C., \& Mathis, J. S. 1989, ApJ, 345, 245  
%\bibitem[Carignan(1985)]{car85}
%	Carignan, C. 1985, ApJS, 58, 107
\bibitem[Carlberg(1992)]{car92}
	Carlberg, R. G. 1992, ApJ, 399, L21
\bibitem[Carlson \& Holtzman(2001)]{ch01}
	Carlson, M. N. \& Holtzman, J. A. 2001, PASP, 113, 1522
\bibitem[Carollo(1999)]{car99}
	Carollo, C. M. 1999, ApJ, 523, 566
%\bibitem[Carollo \ea (1997)]{car97}
%	Carollo, C. M., Stiavelli, M., de Zeeuw, P. T., \& Mack, J. 1997, 
%	AJ, 114, 2366
\bibitem[Carollo \ea (1998)Carollo, Stiavelli, \& Mack]{car98}
	Carollo, C. M., Stiavelli, M., \& Mack, J. 1998, AJ, 116, 68
\bibitem[Carollo \ea (2001)]{car01}
	Carollo, C. M., Stiavelli, M., de Zeeuw, P. T., Seigar, M., \&
	Dejonghe, H. 2001, ApJ, 546, 216
%\bibitem[Courteau, de Jong \& Broeils(1996)]{cou96}
%        Courteau, S., de Jong, R., \& Broeils, A. 1996, ApJL, 457, L73 
\bibitem[Davidge \& Courteau(2002)]{dav02}
	Davidge, T. J. \& Courteau, S. 2002, AJ, 123, 1438
%\bibitem[de Vaucouleurs(1948)]{dev48}
%        de Vaucouleurs, G. 1948, Ann. d'Astrophysique, 11, 247 
%\bibitem[de Vaucouleurs(1963)]{dev63}
%        de Vaucouleurs, G. 1963, ApJ, 137, 720
%\bibitem[de Jong \& Lacey(2000)]{dej00}
%        de Jong, R. S. \& Lacey, C. 2000, ApJ, 545, 781
%\bibitem[de Vaucouleurs \ea (1991)]{dev91}
%        de Vaucouleurs, G., de Vaucouleurs, A., Corwin, H., 
%        Buta, R. J., Paturel, G., \& Fouque, P. 1991, 
%        {\it Third Reference Catalogue of Bright Galaxies}, 
%        New York:Springer-Verlag
\bibitem[Djorgovski \& Meylan(1994)]{djo94}
        Djorgovski, S. \& Meylan, G. 1994, AJ, 108, 1292
\bibitem[Ferrarese \& Merritt(2000)]{fer00}
        Ferrarese, L. \& Merritt, D. 2000, ApJL, 539, 9
\bibitem[Fioc \& Rocca-Volmerange(1997)]{fio97}	
	Fioc, M. \& Rocca-Volmerange, B. 1997, A\&A, 326, 950
%\bibitem[Freedman \ea (1992)]{fre92}
%	Freedman, W. L., Madore, B. F., Hawley, S. L., Horowitz, I. K.,
% 	Mould, J., Navarrete, M., \& Sallmen, S. 1992, ApJ, 396, 80
\bibitem[Friedli (1994)]{frie94}
         Friedli, D. 1994, in Mass-Transfer Induced Activity in Galaxies,
         ed. I. Shlosman (Cambridge: Cambridge Univ.~Press), p. 268
%\bibitem[Friedli \& Benz(1993)]{fri93}
%        Friedli, D. \& Benz, W. 1993, A\&A, 268, 65  
%\bibitem[Fu \ea (2003)Fu, Huang, \& Deng]{fu03}
%        Fu, Y. N., Huang, J. H., \& Deng, Z. G. 2003, MNRAS, 339, 442  
\bibitem[Gebhardt \ea (2000)]{geb00}
        Gebhardt, K. \ea\ 2000, ApJL, 539, 13 
\bibitem[Geha \ea (2002)Geha, Guhathakurta, \& van der Marel]{geh02}
        Geha, M., Guhathakurta, P., \& van der Marel, R. P. 2002,
        AJ, 124, 3073 
\bibitem[Gordon \ea (1999)]{gor99}
        Gordon, K. D., Hanson, M.M., Clayton, G. C., Rieke, G. H.,
	\& Misselt, K. A. 1999, ApJ, 519, 165  
\bibitem[Harris(1996)]{har96}
	Harris, W. E. 1996, AJ, 112, 1487
\bibitem[Harris \ea (2002)]{har02}
	Harris, W. E., Harris, G. L. H., Holland, S. T., \& McLaughlin, D. E.
	2002, AJ, 124, 1435
%\bibitem[Ho \ea (1997)Ho, Filippenko, \& Sargent]{ho97}
%	Ho, L. C., Filippenko, A. V., \& Sargent, W. L. W. 1997, ApJS, 112, 315
\bibitem[Holtzman \ea (1992)]{hol92}
	Holtzman, J. A. \ea\ 1992, AJ, 103, 691
%\bibitem[Gelatt \ea (2000)Gelatt, Hunter, \& Gallagher]{gel00}
%        Gelatt, A. E., Hunter, D. A., \& Gallagher, J. S. 2000, PASP, 113, 142  
%\bibitem[Kormendy \& McClure(1993)]{kor93}
%        Kormendy. J. \& McClure, R. D. 1993, AJ, 105, 1793
%\bibitem[Kormendy \& Bender(1996)]{kor96}
%        Kormendy. J. \& Bender, R. 1996, ApJ, 464, 119
%\bibitem[Krabbe \ea (1995)]{kra95}
%        Krabbe, A. \ea\ 1995, ApJ, 447, L95
%	(Baltimore:STScI)
\bibitem[King(1962)]{kin62}
	King, I. R. 1966, AJ, 67, 471
%\bibitem[King(1966)]{kin66}
%	King, I. R. 1966, AJ, 71, 64 
\bibitem[Kormendy(1985)]{kor85}
	Kormendy, J. 1985, ApJ, 295, 73 
\bibitem[Krist \& Hook (2001)]{kri01}
        Krist, J. \& Hook, R. 2001, The TinyTim User's Guide, Version 6.0
	(Baltimore:STScI)
\bibitem[Kroupa \ea (1993)Kroupa, Tout, \& Gilmore]{kro93}
	Kroupa, P., Tout, C. A., \& Gilmore, G. 1993, MNRAS, 262, 545 
\bibitem[Kundu \& Whitmore(1998)]{kun98}
	Kundu, A. \& Whitmore, B. C. 1998, \aj, 116, 2841 
\bibitem[Larsen(1999)]{lar99}
	Larsen, S. S. 1999, A\&AS, 139, 393 
\bibitem[Larsen(2002)]{lar02}
	Larsen, S. S. 2002, \aj, 124, 1393 
\bibitem[Larsen \ea (2002a)]{lar02b}
	Larsen, S. S., Brodie, J. P., Sarajedini, A., \& Huchra, J. P.
        2002a, \aj, 124, 2615 
\bibitem[Larsen \ea (2002b)]{lar02c}
	Larsen, S. S., Efremov, Y. N., Elmegreen, B. G., Alfaro, E. J., 
	Battinelli, P., Hodge, P. W., \& Richtler, T. 2002b, ApJ, 567, 896 
%\bibitem[Lauer \ea (1995)]{lau95}
%	Lauer, T. R., et al. 1995, \aj, 110, 2622
%\bibitem[Lauer \ea (1998)]{lau98}
%	Lauer, T.~R., Faber, S. M., Ajhar, E. A., Grillmair, C. J., \&
%	Scowen, P. A. 1998, \aj, 116, 2263
%\bibitem[Leitherer \ea (1999)]{lei99}
%	Leitherer, C., et al. 1999, ApJS, 123, 3
\bibitem[Lotz \ea (2001)]{lot01}
	Lotz, J., Telford, R., Ferguson, H. C., Miller, B. W.,
	Stiavelli, M., \& Mack, J. 2001, ApJ, 552, 572
\bibitem[Lucy(1974)]{luc74}
	Lucy L.B. 1974, \aj, 79, 745
\bibitem[Maoz \ea (2001)]{mao01}
	Maoz, D., Barth, A.~J., Ho, L.~C., Sternberg, A., 
	\& Filippenko, A.~V. 2001, \aj, 121, 3048
\bibitem[Matthews \& Gallagher (1997)]{mat97}
	Matthews, L. D. \& Gallagher, J. S., III 1997, \aj, 114, 1899
\bibitem[Matthews \ea (1998)Matthews, van Driel \& Gallagher]{mat98}
	Matthews, L. D., van Driel, W. \& Gallagher, J. S., III 1998, 
	\aj, 116, 2196
\bibitem[Matthews \ea (1999)]{mat99}
	Matthews, L. D. \ea\ 1999, \aj, 118, 208
\bibitem[McLaughlin(2000)]{mcl00}
	McLaughlin, D. E. 2000, ApJ, 539, 618
\bibitem[Mengel \ea (2002)]{men02}
        Mengel, S., Lehnert, M. D., Thatte, N., \& Genzel, R.
        2002, A\&A, 383, 137
%\bibitem[Merritt \& Sellwood(1994)]{mer94}
%        Merritt, D. \& Sellwood, J. A. 1994, ApJ, 425, 551  
\bibitem[Norman \ea (1996)Norman, Sellwood, \& Hasan]{nor96} 
        Norman, C. A., Sellwood, J. A., Hasan, H. 1996, ApJ, 462, 114 
\bibitem[Paturel \ea (1994)Paturel, Bottinelli, \& Gouguenheim]{pat94} 
        Paturel, G., Bottinelli, L., \& Gouguenheim, L. 1994, A\&A, 286, 768
\bibitem[Phillips \ea (1996)]{phi96}
	Phillips, A.C., Illingworth, G. D., MacKenty, J. W., \& Franx, M. 
	1996, \aj, 111, 1566
\bibitem[Press \ea (1992)]{pre92}
	Press, W. H., Teukolsky, S. A., Vetterling, W. T., \& Flannery, B. P.
	1992, {\it Numerical Recipes in C. The art of scientific computing}, 
	Cambridge: Cambridge Univ. Press
%\bibitem[Raha \ea (1991)]{rah91}
%        Raha, A., Sellwood, J.A., James, R., \& Kahn, F.D. 1991, Nat, 352, 411
\bibitem[Rhee(1996)]{rhe96} 
	Rhee, M.-H. 1996, Ph.D. thesis, Univ. Groningen
\bibitem[Richardson(1972)]{ric72} 
	Richardson W.H. 1972, J. Opt. Soc. Am. 62, 55
%\bibitem[Salpeter(1955)]{sal55} 
%	Salpeter, E. E. 1955, ApJ, 121, 161
%\bibitem[Sandage \& Tammann (1990)]{san90}
%	Sandage A., \& Tammann G. 1990, ApJ, 365, 1 
%\bibitem[Sarzi \ea (2002)]{sar02}
%	Sarzi, M. \ea\ 2002, in preparation 
%\bibitem[Schinnerer \ea (2001)Schinnerer, Eckard, \& Tacconi]{sch01}
%	Schinnerer, E., Eckard, A., \& Tacconi, L. J. 2001, ApJ, 549, 254
\bibitem[Schinnerer \ea (2003)Schinnerer, B\"oker, \& Meier]{sch03}
	Schinnerer, E., B\"oker, T., \& Meier, D. S. 2003, ApJ, 591, L115
%\bibitem[Schinnerer \ea (2004)]{sch04}
%	Schinnerer, E., B\"oker, T., Emsellem, E., \& Lisenfeld, U. 2004, 
%	in prep.
%\bibitem[Schlegel \ea (1998)Schlegel, Finkbeiner \& Davis]{sch98}
%         Schlegel, D. J., Finkbeiner, D. P., \& Davis, M. 1998, ApJ, 500, 525
\bibitem[S\'ersic(1968)]{ser68}
        S\'ersic, J.-L. 1968, Atlas de Galaxias Australes 
	(Cordoba: Obs. Astron.)
\bibitem[Shen \& Sellwood (2003)]{shen03}
	Shen, J., \& Sellwood, J. A. 2003, Carnegie Observatories
        Astrophysics Series, Vol. 1: Coevolution of Black Holes and Galaxies,
        ed. L. C. Ho (Pasadena: Carnegie Observatories,
        http://www.ociw.edu/ociw/symposia/series/symposium1/proceedings.html) 
\bibitem[Smith \& Gallagher(2001)]{smi01}
	Smith, L. J. \& Gallagher, J. S. III 2001, MNRAS, 326, 1027
\bibitem[Telesco(1988)]{tel88}
	Telesco, C. M. 1988, ARA\&A, 26, 343
%\bibitem[Tremaine \ea (1975)Tremaine, Ostriker, \& Spitzer]{tre75}
%	Tremaine, S. D., Ostriker, J. P., \& Spitzer, L., Jr. 1975,
%        ApJ, 196, 407
%\bibitem[Tully(1988)]{tul88}
%         Tully, R. B. 1988, Nearby Galaxies Catalog (Cambridge: Cambridge
%	 Univ. Press)
\bibitem[van den Bergh \ea (1991)van den Bergh, Morbey, \& Pazder]{van91}
	van den Bergh, S., Morbey, C., \& Pazder, J. 1991, ApJ, 375, 594 
\bibitem[Walcher \ea (2003)]{wal03}
	Walcher, C. J. \ea\ 2003, Carnegie Observatories Astrophysics 
	Series, Vol. 1: Coevolution of Black Holes and Galaxies, ed. L. C. Ho 
	(Pasadena: Carnegie Observatories,
        http://www.ociw.edu/ociw/symposia/series/symposium1/proceedings.html) 
\bibitem[Whitmore \ea (1999)]{whi99}
	Whitmore, B.~C., Zhang, Q., Leitherer, C., Fall, S.~M., 
	Schweizer, F., \& Miller, B.~W. 1999, \aj, 118, 1551
\bibitem[Zepf \ea (1999)]{zepf99}
        Zepf, S. E., Ashman, K. M., English, J., Freeman, K. C., \&
        Sharples, R. M. 1999, \aj, 118, 752
\end{thebibliography}
\end{document}